\documentclass[12pt]{article}
\usepackage{amsmath}
\usepackage{amsthm}
\usepackage{amssymb}
\usepackage{amsfonts}
\usepackage{times}
\usepackage{macros}

\theoremstyle{plain}

\newtheorem{conj}{Conjecture}

\theoremstyle{remark}

\newcommand{\sst}{\scriptscriptstyle}

\newcommand{\vt}{\vartheta}

\newcommand{\1}{\one}
\newcommand{\2}{\two}

\newcommand{\beq}{\begin{equation}}
\newcommand{\eeq}{\end{equation}}

\newcommand{\pa}{\partial}
\newcommand{\ot}{\otimes}
\newcommand{\ra}{\to}

\newcommand{\SRN}{{\rm N}}

\newcommand{\fr}[2]{{\textstyle \frac{#1}{#2} }}

\newcommand{\bra}{\langle}

\newcommand{\ket}{\rangle}

\newcommand{\Ga}{\Gamma}
\newcommand{\de}{\delta}
\newcommand{\De}{\Delta}
\newcommand{\ep}{\epsilon}

\newcommand{\om}{\omega}

\newcommand{\si}{\sigma}

\newcommand{\vf}{\varphi}

\newcommand{\bk}{{\mathbf k}}

\newcommand{\bn}{\bar{n}}

\newcommand{\bv}{\Bar{v}}

\newcommand{\bx}{\bar{x}}

\newcommand{\CC}{{\mathcal C}}

\newcommand{\CF}{{\mathcal F}}

\newcommand{\CH}{{\mathcal H}}

\renewcommand{\bx}{{\mathbf x}}
\newcommand{\bt}{{\mathbf t}}
\newcommand{\by}{{\mathbf y}}

\newcommand{\CK}{{\mathcal K}}
\newcommand{\CL}{{\mathcal L}}

\newcommand{\CO}{{\mathcal O}}

\newcommand{\CQ}{{\mathcal Q}}

\newcommand{\CS}{{\mathcal S}}
\newcommand{\CT}{{\mathcal T}}

\newcommand{\CV}{{\mathcal V}}

\newcommand{\SA}{{\mathsf A}}
\newcommand{\SB}{{\mathsf B}}
\newcommand{\SC}{{\mathsf C}}
\newcommand{\SD}{{\mathsf D}}

\newcommand{\SG}{{\mathsf G}}

\newcommand{\SK}{{\mathsf K}}

\newcommand{\SM}{{\mathsf M}}

\newcommand{\SO}{{\mathsf O}}

\newcommand{\SQ}{{\mathsf Q}}
\newcommand{\SR}{{\mathsf R}}
\renewcommand{\SS}{{\mathsf S}}
\newcommand{\ST}{{\mathsf T}}
\newcommand{\SU}{{\mathsf U}}

\newcommand{\SW}{{\mathsf W}}

\newcommand{\SY}{{\mathsf Y}}

\newcommand{\sa}{{\mathsf a}}

\renewcommand{\sf}{{\mathsf f}}

\newcommand{\sq}{{\mathsf q}}
\newcommand{\spp}{{\mathsf p}}

\newcommand{\su}{{\mathsf u}}
\newcommand{\sv}{{\mathsf v}}

\newcommand{\sw}{{\mathsf w}}

\newcommand{\sx}{{\mathsf x}}
\newcommand{\sy}{{\mathsf y}}

\newcommand{\BK}{{\mathbb K}}

\newcommand{\0}{{\mathfrak 0}}
\newcommand{\one}{{\mathfrak 1}}
\newcommand{\two}{{\mathfrak 2}}

\newcommand{\BR}{{\mathbb R}}

\newcommand{\BI}{{\mathbb I}}
\newcommand{\BC}{{\mathbb C}}

\newcommand{\BS}{{\mathbb S}}

\newcommand{\BZ}{{\mathbb Z}}

\DeclareMathOperator{\sgn}{sgn}

\newcommand{\rf}[1]{(\ref{#1})}

\newcommand{\aufz}
{\begin{list}{$\bullet$}{\topsep0cm \itemsep0cm \parsep0cm}}
\newcommand{\eaufz}{\end{list}}

\begin{document}


\title{The integrable structure of nonrational conformal field theory}

\author{A. Bytsko$^*$, J. Teschner$^\dagger$}

\address{
$^*$ Steklov Mathematics Institute, Fontanka 27, 191023 St. Petersburg, Russia\\[1ex]
$^\dagger$ DESY Theory,
Notkestr. 85,
22603 Hamburg, Germany}

\begin{abstract}
We define lattice versions of
\end{abstract}

\maketitle

\begin{center}{\it
We dedicate this paper to L.D. Faddeev on the
occasion of his 75th birthday.}
\end{center}

\section{Introduction}

The main question which motivated this work is the following:
How do conformal field theories look like if studied from the point of
view of a possibly existing {\it integrable structure}?
There are many quantum-field theoretical models of high interest for string theory and
condensed matter physics which are expected to have
conformal invariance, but not enough chiral symmetry to make a solution in terms
of standard methods of conformal field theory look realistic. An interesting class of examples are
nonlinear sigma models with targets being super-groups, which have recently
attracted considerable interest both from string theory and condensed matter
physics. Some of these theories are expected to be integrable. It therefore seems
reasonable to expect that methods from the theory of integrable models
can be used to understand the spectrum of these theories.

Such a program immediately faces an obstacle: Up to now it seemed that key
features of conformal field theories like the factorization into left- and
right-moving degrees of freedom are very hard to see with the help of the
integrable structure. Using the traditional approaches
based on the Bethe ansatz one usually has to go a rather long way until some of the
features of conformal invariance become visible. We therefore looked for
a simple, but prototypical example where we can improve on this state of
affairs. The main point we want to illustrate with the example of
Liouville theory is the following: The factorization into left- and
right-movers can be made manifest in a very transparent way already on the
level of an integrable lattice regularization of a conformal field theory.

The framework in which this turns out to be the case combines the use
of Baxter's Q-operators with the Separation of Variables technique of
Sklyanin \cite{Sk1,Sk2,Sk3}.
In the cases under consideration we will explicitly construct
Q-operators $\SQ^+(u)$ and $\SQ^-(u)$ which contain the conserved charges
of left- and right-moving degrees of freedom, respectively. Within the
Separation of Variables framework one may then represent
an eigenstate of $\SQ^+(u)$ and $\SQ^-(u)$ in terms of a wave-function
constructed directly out of the corresponding eigenvalues
$q^+(u)$ and $q^-(u)$. The combination of these two ingredients
yields
a quantum version of the B\"acklund transformation from
Liouville theory to free field theory, making the factorization
into left- and right-moving degrees of freedom transparent.

It also seems promising to view the integrable structure of conformal
field theories as a useful starting point for the study of massive integrable
models. One may expect that the integrable structure
"deforms smoothly" from the massless to the massive cases, but is simpler to
study in the massless limits.
This point of view was developed in particular in the beautiful series
of works \cite{BLZ,BLZ2}, where conformal field theories with central charge $c<1$ were studied.
One of our aims here is to study the counterpart of this theory
for $c>1$. The constructions from \cite{BLZ}
no longer work in this case due to more severe ultraviolet problems.
We will use an integrable lattice regularization to control such problems.
This will also allow us
study the Sinh-Gordon model, Liouville theory and quantum KdV theory in a
uniform framework. We will observe that
key objects of the integrable structure like the Baxter Q-operators are
indeed related to each other by certain parametric limits.

The example chosen, Liouville theory, is of considerable interest in its
own right. It has attracted a lot of attention for more than 25 years now
due to its connections with noncritical string theory and two-dimensional quantum
gravity (see \cite{DG,GM} for reviews and references), 
as an example for interesting non-rational conformal field 
theories \cite{TR,TCFT},
and due to its relations to the (quantized) Teichm\"uller spaces
of Riemann surfaces \cite{TT,T07}.

In the study of Liouville theory, the most popular approach so far was based
on its conformal symmetry, leading to a complete solution in the sense of the
Belavin-Polyakov-Zamolodchikov bootstrap approach \cite{BPZ}, see
\cite{CT,GN,DO,ZZ,PT99,TL} for some 
key steps in this program, and \cite{TR} for a
more complete list of references. Understanding Liouville theory from the point
of view of its integrable structure has also attracted considerable interest
in the past, going back to \cite{FT86}, and more recently being developed in \cite{FKV,FK}.
This approach has also lead to beautiful results, see in particular \cite{FK}.

What seemed somewhat unsatisfactory, however, was the lack of
results that can be directly compared with the conformal field theory approach.
It is the second main aim of this paper to re-derive the so-called
reflection amplitude of Liouville theory with the help of its integrable
structure. The formula for this quantity had been conjectured in
\cite{DO,ZZ}. A derivation for these conjectures was subsequently given in
\cite{TL}. Here we are going to re-derive this result in a completely
different way, entirely based on the integrable structure of Liouville theory.

However, we feel that the interplay between conformal and integrable
structures is still not completely understood. It seems particularly
important to integrate the lattice Virasoro algebra \cite{FV93} into
the picture and to clarify the relations with the beautiful work 
\cite{BMS} where closely related models of statistical 
mechanics were studied.
What we do hope, however, is that this paper lays some useful
groundwork which will ultimately lead to a better understanding of this
important subject. 

This paper is intended to give a reasonably concise overview over the main
constructions, ideas and results of our work.
It is not self-contained. In order to make
the verification of our claims possible, we either give sketches of
the proofs or indicate references where similar arguments can be found.
A more detailed presentation is in preparation.

\vspace*{1mm}
{\par\small
{\em Note on notations:} In order to distinguish objects 
associated to the three different models of interest, we shall
sometimes use subscripts like $\SO_{\rm\sst ShG}$, 
$\SO_{\rm\sst Liou}$ or $\SO_{\rm\sst KdV}$. 
However, to unload the notation we shall omit 
these subscripts whenever it is clear from the
context which model is considered.
}

\vspace*{1mm}
{\par\small
{\em Acknowledgements.} We would like to
thank V. Bazhanov, A. Bobenko, R. Kashaev
for stimulating discussions. Many thanks also
to L. Faddeev and V. Bazhanov for comments on the draft.

J.T. gratefully acknowledges support 
from the EC by the Marie Curie Excellence
Grant MEXT-CT-2006-042695.
A.B. was supported in part by the RFBR under grants
 07-02-92166 and 08-01-00638.}

\section{Definition of the lattice models}

\setcounter{equation}{0}

The aim of this section is to define three lattice models, 
corresponding to
the Sinh-Gordon model, Liouville theory and the 
scalar free field theory, respectively. 
Anticipating discussions of its integrable structure we will refer to
the scalar free field 
theory as KdV theory below.

\subsection{Lattice discretization}

The classical counterparts of the models in question are dynamical systems
whose degrees of freedom are described by
the field $\phi(x,t)$ defined for
$(x,t)\in[0,R]\times \BR$ with periodic boundary conditions $\phi(x+R,t)=\phi(x,t)$.
The dynamics of these models may be described in the Hamiltonian form in terms of variables
$\phi(x,t)$, $\Pi(x,t)$, the Poisson brackets being
\[
\{\,\Pi(x,t)\,,\,\phi(x',t)\,\}\,\,=\,2\pi\,\de(x-x')\,.
\]
The time-evolution of an arbitrary observable $O(t)$ is then given as
\[
\pa_tO(t)\,=\,\{\,H\,,\,O(t)\,\}\,,
\]
with Hamiltonian $H$ being defined as
\begin{equation}
H\,=\,\int_0^{R}\! \frac{dx}{4\pi}\,h(x)\,,\qquad
\begin{aligned}
&h_{\rm\sst ShG}=\Pi^2+(\pa_x \phi)^2+8\pi\mu \cosh(2b\phi)\,,\\
&h_{\rm\sst Liou}=\Pi^2+(\pa_x \phi)^2+4\pi\mu e^{-2b\phi}\,,\\
&h_{\rm\sst KdV}=\Pi^2+(\pa_x \phi)^2\,.
\end{aligned}
\end{equation}
In order to regularize the ultraviolet divergencies that arise in the quantization of these models
we will pass to integrable lattice discretizations.
First discretize the field variables according to the standard recipe
\begin{equation*}\label{cont}
 \phi_n \equiv \phi(n\Delta) \,, \quad
 \Pi_n \equiv \De\Pi(n\Delta) \,,
\end{equation*}
where $\Delta=R/\SRN$ is the lattice spacing.
Quantization is then canonical: The variables $\Phi_n$, $\Pi_n$, 
$n\in\BZ/\SRN\BZ$ are henceforth
considered as operators with commutation relations
\begin{equation}
{[}\,\phi_n\,,\,\Pi_m\,{]}\,=\,2\pi i\de_{n,m}\,,
\end{equation}
that can be realized in  the usual way on the Hilbert space $\CH\equiv(L^2(\BR))^{\ot \SRN}$.
As another convenient set of variables let us introduce the operators $\sf_{k}$ defined as
\begin{equation} \sf_{2n}\,\equiv\,e^{-2b\phi_n}\,,\qquad
\sf_{2n-1}\,\equiv\,e^{\frac{b}{2}(\Pi_n+\Pi_{n-1}-2\phi_n-2\phi_{n-1})}\,.
\end{equation}
This change of variables is invertible for $\SRN\equiv 2L+1$ 
odd. We will therefore restrict our attention to this case in
the following. The variables $\sf_n$ satisfy the
algebraic relations
\begin{equation}
\sf_{2n\pm 1}\,\sf_{2n}\,=\,q^2\,\sf_{2n}\,\sf_{2n\pm 1}\,,\quad q=e^{\pi i b^2}\,,\qquad
\sf_{n}\,\sf_{n+m}\,=\,\sf_{n+m}\,\sf_{n}\;\;{\rm for}\;\;m\geq 2\,.
\end{equation}
These operators turn out to represent the initial data 
for time evolution in a particularly convenient way, as we are
going to discuss next.

\subsection{Lattice dynamics}

A beautiful way to define a suitable dynamics in these lattice models was proposed by
Faddeev and Volkov in \cite{FV94}. This approach was adapted to the lattice Liouville model in \cite{FKV}.
Space-time is replaced by the cylindric lattice
\[
\CL\,\equiv\,\big\{\,(\nu,\tau)\,,\,\nu\in\BZ/\SRN\BZ\,,\,\tau\in\BZ\,,\,\nu+\tau={\rm even}\,\big\}\,.
\]
The condition that $\nu+\tau$ is even means that the lattice is rhombic: The lattice points closest
to $(\nu,\tau)$ are $(\nu\pm 1,\tau+1)$ and $(\nu\pm 1,\tau-1)$.
We identify the variables $\sf_n$ with the initial values of a discrete "field" $\sf_{\nu,\tau}$ as
\[ \sf_{2r,0}\,\equiv\,\sf_{2r}\,,\qquad
\sf_{2r-1,1}\,\equiv\,\sf_{2r-1}\,.
\]
One may then extend the definition recursively to all $(\nu,\tau)\in\CL$ by
\begin{equation}\label{Hirota}
{\sf}_{\nu,\tau+1}\,\equiv\,\sf_{\nu,\tau-1}^{-\frac{1}{2}}\cdot g_\kappa^{}\big(\sf_{\nu-1,\tau}^{}\big)g_\kappa^{}\big(\sf_{\nu+1,\tau}^{}\big)\cdot\sf_{\nu,\tau-1}^{-\frac{1}{2}}\,,\\
\end{equation}
with functions $g$ defined respectively by
\begin{equation}\label{gkappadef}\begin{aligned}
& g_\kappa(z)\,=\,\frac{\kappa^2+z}{1+\kappa^2 z}  \\
& g_\kappa(z)\,=\,\frac{z}{1+\kappa^2 z}\\
& g_\kappa(z)\,=\,z
\end{aligned}
\quad\begin{aligned}
&\text{for the Sinh-Gordon model,}\\[.75ex]
&\text{for Liouville theory,}
\\[.75ex]
&\text{for KdV theory.\,}
\end{aligned}
\end{equation}
where $\kappa$ plays the role of a scale-parameter of the theory.
In the massive case it can 
be identified with a certain function of the physical mass \cite{T}.
We refer to \cite{FV94} for a nice discussion of the relation between the lattice evolution equation \rf{Hirota} and the classical Hirota
equation, explaining in particular how to recover
the Sinh-Gordon equation in the classical continuum limit.

In order to construct the  unitary operators $\SU$ that generate the time evolution above
let us, following \cite{FKV} closely, introduce the special functions 
$w_b(x)$ and $\vf(x)$ which are defined as
\begin{equation}\label{wphidef}
w_b(x)\,=\,\frac{\zeta\,e^{\frac{\pi i }{2}x^2}}{\vf(x)}
\,,
\qquad\vf(x)\,=\,\exp\left(\int_{\BR+i0}\frac{dt}{4t}\;\frac{e^{-2itx}}{\sinh (bt)\sinh (b ^{-1}t)}\right)\,,
\end{equation}
where $\zeta=e^{\frac{\pi i}{24}(b^2+b^{-2})}$.
The special function $\vf(x)$ has been introduced
in a related context in \cite{F95}. All the relevant properties
(zeros, poles, asymptotic behavior, functional relations)
can be found in \cite{Vo05,BT,BMS}.
Out of these functions let us construct
\begin{equation}\label{wvdef}\begin{aligned}
& G_{v}(e^{2\pi b x})\,=\,{w_b(\fr{v}{2}+x)}{w_b(\fr{v}{2}-x)}\,\\[.5ex]
& G_{v}(e^{2\pi b x})\,=\,\zeta^{-1}\,
e^{-i\frac{\pi}{2}(x+\frac{v}{2})^{2}}
\,w_b\big(\fr{v}{2}- x\big)\\[.5ex]
& G_{v}(e^{2\pi b x})\,=\,\zeta^{- 2}\,
e^{- i\frac{\pi}{2}(x+\frac{v}{2})^{2}}
\,e^{- i\frac{\pi}{2}(x-\frac{v}{2})^{2}}
\end{aligned}\quad\begin{aligned}
&\text{for the Sinh-Gordon model,}\\[.5ex]
&\text{for Liouville theory,}
\\[.5ex]
&\text{for KdV theory.\,}
\end{aligned}
\end{equation}
Let us then consider the operator $\SU$, defined as
\begin{equation}
\SU\,=\,\prod_{n=1}^{\SRN}G_{2s}(\sf_{2n})\cdot\SU_0\cdot\prod_{r=1}^{\SRN}G_{2s}(\sf_{2n-1})\,,
\end{equation}
where $\SU_0$ is the parity operator that acts as $\SU_0^{}\cdot\sf_k^{}=\sf^{-1}_k\cdot\SU_0^{}$.
The functions $G_{2s}(z)$ satisfy the functional relations
\begin{equation}\label{funrel}
G_{2s}(qz)\,/\,G_{2s}(q^{-1}z)\,=\,g_\kappa(z)\,\quad{\rm if}\;\;\kappa=e^{-\pi b s}\,,
\end{equation}
where $G_v$ and $g_\kappa$ are chosen from \rf{wvdef} and \rf{gkappadef} according to the case at hand.
It easily follows from \rf{funrel} that $\SU$ is indeed the the generator of the time-evolution \rf{Hirota},
\begin{equation}
\sf_{\nu,\tau+1}\,=\,\SU^{-1}\cdot \sf_{\nu,\tau-1}\cdot\SU\,.
\end{equation}
One of our tasks is to exhibit the integrability of this discrete time evolution.

\subsection{Fock space representation}\label{focksec}

Classically the Hamiltonian density of KdV theory is the one of a free field theory. The correspondence with
free field theory becomes manifest in the lattice model if we introduce 
lattice analogs of the fields $e^{b(\pa_t\pm \pa_x)\phi}$ as follows 
\cite{Ge,Vo92}
\begin{equation}\begin{aligned}
&\sw_n^+\,=\,q\,\sf_{2n+1}^{}\sf_{2n+2}^{-1}
\,,\\
&{\sw}_n^-\,=\,q\,\sf_{2n+1}^{}\sf_{2n}^{-1}
\,,\end{aligned}
\begin{aligned}
&\qquad \sw_{\nu,\tau}^+\,\equiv\,q\,\sf_{\nu,\tau}^{\phantom{-1}}\,
\sf_{\nu+1,\tau-1}^{-1}\,,\\
&\qquad \sw_{\nu,\tau}^-\,\equiv\,q\,\sf_{\nu,\tau}^{\phantom{-1}}\,
\sf_{\nu-1,\tau-1}^{-1}\,.
\end{aligned}\end{equation}
Note that the operators $\sw_n^+$, $\sw_n^-$ satisfy the following commutation relations:
\begin{align}\label{+-comm}
& \sw_n^+\sw_m^-\,=\,\sw_m^-\sw_n^+\,,
\quad \begin{aligned}&\sw_n^+\sw_m^+\,=\,\om_{nm}^{} \sw_m^+\sw_n^+,\\
&\sw_n^-\sw_m^-\,=\,\om_{nm}^{-1} \sw_m^-\sw_n^-,
\end{aligned}\quad
\om_{nm}\equiv\left\{\begin{aligned} q^{{2}\sgn(m-n)} \;\,&
{\rm if}\;\,
|n-m|=1,\\ 1
 \;\,&
{\rm if}\;\,
|n-m|\neq 1.\end{aligned}\right.
\end{align}
The evolution generated by the operator $\SU_{\rm\sst KdV}$ is represented in these variables as
\begin{equation}\label{w-dyn}
\sw_{\nu,\tau+1}^+\,=\,\sw_{\nu-1,\tau}^+\,,\qquad
\sw_{\nu,\tau+1}^-\,=\,\sw_{\nu+1,\tau}^-\,.
\end{equation}
This means that that the variables $\sw_n^+$ and $\sw_n^-$ represent the right and the left-moving degrees
of freedom respectively.


We will sometimes 
use an alternative representation for the Hilbert space 
$\CH$ which not only makes
the  chiral factorization into left- and right-moving degrees 
manifest for KdV-theory, but will
also be used in the discussion of Liouville theory.
Keeping in mind $\SRN=2L+1$ let
\begin{equation}\begin{aligned}
&\spp_\0^{}\,=\,\frac{1}{2\pi b\SRN}\sum_{n=-L}^{L}\,\log\sw_n^{\pm}\,,\qquad
\sq_\0^{}\,=\,\frac{1}{2\pi}\sum_{n=1}^\SRN\phi_n\,,\\
&\sa_k^{\pm}\,\equiv\,\frac{1}{2\pi b}
\sum_{n=-L}^{L}\,e^{2i\frac{\pi}{\SRN}nk}\,
\left( {\log \sw_n^\pm}-2\pi b\spp_\0\right)\,.
\end{aligned}\end{equation}
We have the following commutation relations,
\begin{equation}
\begin{aligned}
& {[}\,\sa_n^{+}\,,\,\sa_m^{-}\,]\,=\,0\,,\qquad
[\,\sa_n^{\pm}\,,\,\sa_m^{\pm}\,]\,=\,\pm\de_{n+m,0}\,
\frac{\sin 2\rho n}{\rho}\,,\qquad \rho\,\equiv\,
\frac{\pi}{N}\\
& [\,\spp_\0^{}\,,\,\sq_\0^{}\,]\,=\,(2\pi i)^{-1}\,,\qquad
[\,\sq_\0^{}\,,\,\sa_n^{\pm}\,]\,=\,0\,,\qquad
[\,\spp_\0^{}\,,\,\sa_n^{\pm}\,]\,=\,0\,.
\end{aligned}\end{equation}
Let
$\CF^{\pm}$ be the Fock spaces generated by the
harmonic oscillators $(\sa_n^\pm,\sa_{-n}^\pm)$ for $n\neq 0$, respectively.
There are representations for the
Hilbert space $\CH_{\rm\sst SG}$ in which either $\spp_\0$ or $\sq_\0$
are represented as multiplication operators,
\begin{equation}\label{CHspecK}\begin{aligned}
\CH_{\rm\sst SG}\,&
\simeq\,\CH_{\rm\sst Fock}\,\equiv\,
\int_{-\infty}^{\infty}dp\;\,\CF_p^+\ot\CF_p^-\,,\\
& \simeq\,\CH_{\rm\sst Schr}\,\equiv\,
\int_{-\infty}^{\infty}d\phi_\0\;\,\CF_{\phi_\0}^+\ot\CF_{\phi_\0}^-\,,
\end{aligned}\qquad
\begin{aligned}& \spp_\0\,(\CF_p^+\ot\CF_p^-)\,=\,p\,(\CF_p^+\ot\CF_p^-)
\\[2.5ex]
& \sq_\0\,(\CF_{\phi_\0}^+\ot\CF_{\phi_\0}^-)\,=\,\phi_0\,
(\CF_{\phi_\0}^+\ot\CF_{\phi_\0}^-)\,.
\end{aligned}\end{equation}
These representations $\CH_{\rm\sst Fock}$ and $\CH_{\rm\sst Schr}$
for $\CH$ will be called the Fock and the (zero mode) 
Schr\"odinger representation,
respectively.

\section{Integrability}

\setcounter{equation}{0}

In order to  exhibit the integrability of the discrete time evolutions introduced in the
previous section one needs to construct mutually commutative families $\CQ$ of self-adjoint operators $\ST$ such that
\begin{equation}
\begin{aligned}
& {\rm (A)}\quad[\,\ST\,,\,\ST'\,]\,=\,0,\\
& {\rm (B)}\quad[\,\ST\,,\,\SU\,]\,=\,0,\\
& {\rm (C)}\quad{\rm if}\;\;[\,\ST\,,\,\SO\,]\,=\,0,
\end{aligned}\;\;
\begin{aligned}
& \forall\, \ST,\ST'\in\CQ\,,\\
& \forall\,\ST\in\CQ\,,\\
&\forall \,\ST \in \CQ, \;\;{\rm then}\;\;\SO=\SO(\CQ).
\end{aligned}\end{equation}
Within the framework of the quantum inverse scattering method one may conveniently define the family $\CQ$ in
terms of one-parameter families  $\ST(u)$ and $\SQ(v)$ of operators that are mutually commuting for
arbitrary values of the spectral parameters $u$ and $v$, and which satisfy a functional relation of
the form
\begin{equation}
\ST(u)\SQ(u)\,=\,a(u)\SQ(u-ib)+d(u)\SQ(u+ib)\,,
\end{equation}
with $a(u)$ and $d(u)$ being certain model-dependent coefficient functions. The generator of lattice time
evolution will be constructed from the specialization of the Q-operators to certain values of the spectral
parameter $u$, making the integrability of the evolution manifest.

\subsection{T-operators}

The definition of T-operators for the models in question is standard. It is of the general form
\begin{equation}
\ST^{}(u)\,=\,{\rm tr}_{\BC^2}^{}M(u)\,,\qquad M(u)\,\equiv\,
L_N^{}(u)L_{N-1}^{}(u)\dots L_1^{}(u)\,.
\end{equation}
In the following subsection we will describe
possible  choices for the Lax-matrices $L_n(u)$ for the models of interest.

\subsubsection{Sinh-Gordon model}

For future use let us note that the
L-operator of lattice Sinh-Gordon model \cite{FST,IK,Sk83} can be written as
\begin{equation}\label{LL}
L_n(u)\,\equiv\,L_n(\mu,\bar\mu)\,=\,\left(
\begin{matrix} \su_n+{\mu}\bar\mu^{-1} \sv_n\su_n\sv_n &
\mu \sv_n^{}+\bar\mu^{-1} \sv_n^{-1} \\
\mu \sv_n^{-1}+\bar\mu^{-1} \sv_n^{}
& \su_n^{-1}+{\mu}\bar\mu^{-1} \sv_n^{-1}\su_n^{-1}\sv_n^{-1}
\end{matrix}
\right)\,,
\end{equation}
where we have used the notations
\[
\su_n\,=\, e^{\frac{b}{2}\Pi_n}\,,\quad
\sv_n\,=\,e^{-b\phi_n}\,,\quad
\mu\equiv   -i e^{\pi b(u - s)}\,,  \quad \bar{\mu}
\equiv  -i e^{\pi b(u + s)}\,.
\]
The key point about the definition \rf{LL} is the fact that
the commutation relations for the matrix elements of $L_n(u)$
can be written in the Yang-Baxter form
\begin{equation}\label{RLLLLR}
R_{12}^{}(u-v)L_{1n}^{}(u)L_{2n}^{}(v)\,=\,L_{2n}^{}(u)L_{1n}^{}(v)R_{12}^{}(u-v)\,,
\end{equation}
where the $4\times 4$-matrix $R_{12}(u-v)$ is
\begin{equation}\label{R12def}
R(u)\,=\,\left(
\begin{matrix} \sinh\pi b(u+ib) & & & \\
&\sinh\pi b u & i\sin\pi b^2 & \\
& i\sin\pi b^2 & \sinh\pi b u & \\
& & &\sinh\pi b(u+ib)
\end{matrix}\right)\,.
\end{equation}
This implies as usual that the one-parameter family of operators $\ST(u)$
is mutually commutative,
$
[\ST(u),\ST(v)]=0.
$

\subsubsection{Liouville theory}

Faddeev-Tirkkonen \cite{FT} proposed the following L-matrix for the lattice Liouville model,
\begin{align}
&L^{^+}_{{\rm\sst Liou},n}(\mu,\bar\mu)\,=\,\left(
\begin{matrix} \su_n+ \mu\bar\mu^{-1}\sv_n\su_n\sv_n &
\mu \sv_n^{} \\
\mu \sv_n^{-1}+\bar\mu^{-1} \sv_n^{}
& \su_n^{-1}
\end{matrix}
\right)\,.
\end{align}
This L-matrix can be obtained from $L_{{\rm\sst ShG},n}(\mu,\bar\mu)$ in the limit
\begin{equation}\label{LimL1}
L^{^+}_{{\sst\rm Liou},n}(\mu,\bar{\mu})\,\equiv\,\lim_{s\ra\infty}e^{-\frac{\pi}{2} bs\si_3}\,\su_n^{\frac{s}{ib}}\cdot
L_{{\sst\rm ShG},n}^{}(\mu,e^{+2\pi bs}\bar{\mu})\cdot\su_n^{-\frac{s}{ib}}\,e^{+\frac{\pi}{2} bs\si_3}\,,
\end{equation}
and it also satisfies \rf{RLLLLR}. However,
it is easy to see that the corresponding transfer matrix
\begin{equation}\label{LiouT}
\ST^{^+}(u)\,=\,{\rm tr}_{\BC^2}^{}(L^{^+}_\SRN(u)\cdots L^{^+}_{1}(u))
\end{equation}
generates only $L+1$ commuting operators if we have $N=2L+1$ degrees of freedom. $\ST^{^+}(u)$ alone will therefore
{\it not} generate sufficiently many conserved quantities.

Fortunately there exist a second reasonable limit
\begin{equation}\label{LimL2}
L_{{\sst\rm Liou},n}^{^-}(\mu,\bar{\mu})\,\equiv\,\lim_{s\ra\infty}e^{+\frac{\pi}{2} bs\si_3}\,\su_n^{\frac{s}{ib}}\cdot
L_{{\sst\rm SG},n}^{}(e^{-2\pi bs}\mu,\bar{\mu})\cdot\su_n^{-\frac{s}{ib}}\,e^{-\frac{\pi}{2} bs\si_3}\,,
\end{equation}
which leads to yet another solution to \rf{RLLLLR}, namely
\begin{align}
&L^{^-}_{{\rm\sst Liou},n} (\mu,\bar\mu)\,=\,\left(
\begin{matrix} \su_n+ \mu\bar\mu^{-1}\sv_n\su_n\sv_n &
\mu \sv_n^{}+\bar\mu^{-1}\sv_n^{-1} \\
\bar\mu^{-1} \sv_n
& \su_n^{-1}
\end{matrix}
\right)\,.
\end{align}
The mutual commutativity of $\ST^{^+}(u)$ and
$\ST^{^-}(v)$ for all $u,v$ follows by standard arguments from the
commutation relations
\begin{equation}
R_{12}'(u-v)\,L^{^+}_{1}(u)\, L_{2}^{^-}(v)\,=\, L_{2}^{^-}(v)\,L_{1}^{^+}(u)\,R_{12}'(u-v)\,,
\end{equation}
where
\begin{equation}
R_{12}'(u)=\left(
\begin{matrix} e^{\pi b(u+ib)} & & & \\
& e^{\pi b u} & 0 & \\
& i\sin\pi b^2 & e^{\pi b u} & \\
& & & e^{\pi b(u+ib)}
\end{matrix}\right).
\end{equation}
We will later show that the splitting of the transfer matrix $\ST(u)$ into
$\ST_{\rm\sst Liou}^{^+}(u)$ and
$\ST_{\rm\sst Liou}^{^-}(v)$ reflects the chiral factorization of Liouville theory into
left- and right-moving degrees of freedom.

\subsubsection{KdV theory}

The operators $\ST^\pm(u)$ for lattice KdV theory 
can finally be constructed from the Lax-matrices \cite{Ge,Vo92}
\begin{align*}
L_{n}^{^+}(\mu)\,\equiv\,
\left(\begin{matrix}
\su_n & \mu \,\sv_n\\
\mu\,\sv^{-1}_n & \su^{-1}_n
\end{matrix}\right)\,,\quad
L_n^{^-}(\bar\mu)\,\equiv\,
\left(\begin{matrix}
\su_n & \bar\mu^{-1} \,\sv_n^{-1}\\
\bar\mu^{-1} \,\sv_n & \su^{-1}_n
\end{matrix}\right)\,.
\end{align*}
These L-matrices also satisfy \rf{RLLLLR} and
can be obtained \cite{Vo92} from 
$L_{{\rm\sst ShG},n}(u)$  and $L_{{\rm\sst Liou},n}^\pm(u)$
by certain limiting procedures similar to \rf{LimL1},\rf{LimL2}.

It was shown in Subsection \ref{focksec} that the decoupling of the free field dynamics into
right- and left-moving degrees of freedom becomes manifest in the lattice model in terms of
the variables $\sw_n^+$ and $\sw_n^-$. It is possible to show \cite{Vo92} 
that the transfer matrices
$\ST^\ep(u)$,  $\ep=\pm$,
can be represented as a polynomial in the
variables $\sw_n^\ep$ which is independent of $\sw_n^{-\ep}$.

\subsection{Construction of Q-operators}

Algebraic constructions of Q-operators have previously been given in \cite{Vo97} for the KdV model\footnote{More precisely its
chiral half, as will become clear later.} and for the lattice Liouville theory \cite{FKV,Ka01}. It has to be observed, however,
that only the Q-operator related to the T-operator $\ST^+_{\rm\sst Liou}$ by means of
a Baxter-type equation was considered in \cite{FKV,Ka01}.
We observed in the previous subsection that the T-operator $\ST^+_{\rm\sst Liou}$ does not generate sufficiently many conserved
quantities. This suggests that we need a second Q-operator $\SQ^-_{\rm\sst Liou}$ related to $\ST^-_{\rm\sst Liou}$
by a Baxter-type relation in order to complete the proof of the integrability of the lattice Liouville model in the sense formulated above.

We will in the following give a uniform construction of Q-operators for all the models in question.
For our purposes it will be most convenient to represent the Q-operators as integral operators with
explicitly specified integral kernels. 
This facilitates the derivation of the analytic properties of the
Q-operators, as first done in \cite{BT} for the 
Sinh-Gordon model, considerably.


\subsubsection{Representations as integral operators}

In order to represent the Q-operators as integral operators it will be convenient to use the
representation where the operators
$\su_r$ and $\sv_r$ are represented as
\begin{equation}
\su_n\,=\,e^{\pi b(2\sx_n-\spp_n)}\,\qquad
\sv_n\,=\,e^{\pi b\spp_n}\,,
\end{equation}
with $\sx_n$, $\spp_n$ being realized
on wave-functions $\Psi(\bx)$, $\bx=(x_1,\dots,x_\SRN)$ as
\[
\sx_n\cdot\Psi(\bx)\,=\,x_n\Psi(\bx)\,,
\qquad\spp_n\cdot\Psi(\bx)\,=\,\frac{1}{2\pi i}\frac{\pa}{\pa x_n}\Psi(\bx)\,.
\]
Out of the special function $w_b(x)$ let us  form a few useful combinations:
\begin{equation}\begin{aligned}
& \overline{W}_{v-i\eta}(x)\,=\,
 \frac{w_b(x-\frac{v}{2})}{w_b(x+\frac{v}{2})}\,,\\
& W_{i\eta+v}^{+\rm\sst ShG}(x)\,=\,\big(W_{i\eta-v}^{-\rm\sst ShG}(x))^{-1}
\,=\,\frac{w_b(x+\fr{v}{2})}{w_b(x-\fr{v}{2})}\,,\\
& W_{i\eta+v}^{+\rm\sst Liou}(x)\,=\,
\big(W_{i\eta-v}^{-\rm\sst Liou}(x)\big)^{-1}\,=\,
\frac{\zeta^{-1}e^{- i\frac{\pi}{2}(x+\frac{v}{2})^{2}}}
{ w_b\big(x-\fr{v}{2}\big)}\,,\\
& W_{i\eta+v}^{+\rm\sst KdV}(x)\,=\,
\big(W_{i\eta-v}^{-\rm\sst KdV}(x)\big)^{-1}\,=\,
\frac{\zeta^{-1}e^{- i\frac{\pi}{2}(x+\frac{v}{2})^{2}}}
{\zeta^{+1}e^{+i\frac{\pi}{2}(x-\frac{v}{2})^{2}}}\,,
\end{aligned}\qquad
\eta\equiv\frac{1}{2}(b+b^{-1})\,.
\end{equation}
From the known asymptotic properties of the function $w_b(x)$ it is easily found that
$W_{v}^{\pm\rm\sst Liou}$ and $W_{v}^{\pm\rm\sst KdV}$ can be obtained from $W_{v}^{\pm\rm\sst ShG}$
by taking suitable limits.

The Q-operators may then be constructed in the following general form:
\begin{equation}
\SQ^+(u)\,=\,\SY_{\infty}^{-1}\cdot\SY^+_{}(u)\,,\qquad
\SQ^-(u)\,=\,\SY^-_{}(u)\cdot\SY_{-\infty}^{-1}\,,
\end{equation}
where $\SY^\ep(u)$ can be represented as integral operators with kernels
\begin{align}\label{Ykern}
\bra  & \,{\mathbf x}'\,|\,\SY^+(u)\,|\,{\mathbf x}\,\ket\,=\,
\prod_{n=1}^\SRN
\overline{W}_{u-s}^{}(x_{n}'-x_n^{})W_{u+s}^{+}(x_{n-1}'+x_n^{})\,,\\
\bra  & \,{\mathbf x}'\,|\,\SY^-(u)\,|\,{\mathbf x}\,\ket\,=\,
\prod_{n=1}^\SRN W_{u-s}^-(x_{n-1}'+x_n^{})\overline{W}_{u+s}^{}(x_n'-x_{n}^{})
\,,
\label{Ybarkern}\end{align}
whereas the operators $\SY_{\pm\infty}$ have the distributional kernels
\begin{equation}
\bra   \,{\mathbf x}'\,|\,\SY_{\pm\infty}\,|\,{\mathbf x}\,\ket\,=\,
\prod_{n=1}^\SRN e^{\mp 2\pi i x_n'(x_n+x_{n+1})}\,.
\end{equation}
The expressions for the kernel of the operators $\SY^{\ep}(u)$ 
are very similar to the remarkable
factorized expressions for the matrix elements of Q-operators  
found in \cite{BS} for models with related quantum algebraic structures. 
We will present a systematic procedure to derive
such factorized expressions for a certain class of 
models in \cite{BT09}.

The mutual commutativity of T- and Q-operators,
\begin{equation}
\big[\,\SQ^\ep(u)\,,\,\SQ^{\ep'}(v)\,\big]\,=\,0\,,\qquad
\big[\,\SQ^\ep(u)\,,\,\ST^{\ep'}(v)\,\big]\,=\,0\,,\qquad
\ep,\ep'=\pm\,,
\end{equation}
can be shown either along the lines of \cite{BS,PG,BT}
from the star-triangle relation
satisfied by the function $W_u(x)$ \cite{Ka00,Vo05,BT,BMS}
\footnote{The papers \cite{Ka00,Vo05} derive integral 
identities which
can be rewritten in the form of the 
star-triangle relation \cite{BT,BMS}. 
An 
elegant proof can be given by using arguments similar to 
\cite{Ba08} from the
Yang-Baxter equation satisfied by the corresponding R-matrix},
or more elegantly  by writing
the Q-operators as traces of generalized monodromy matrices over q-oscillator
type representations in auxilliary space \cite{BT09}, 
similar to the constructions of Q-operators in \cite{BLZ}.

\subsection{Proof of integrability}

The key observation proving the integrability of the models is the fact that
\begin{equation}\label{UfromQ}
\boxed{\quad \SU\,=\,\SU^+_{}\cdot\SU^-_{}\qquad\SU^+_{}\,=\,\SQ^+_{}(s_+)
\qquad
\SU^-_{}\,=\,(\SQ^-_{}(s_-))^{-1}\quad}
\end{equation}
where we have introduced the notations $s_+=s-i\eta$, $s_-=-s-i\eta$ for
convenience.
The operators $\SU^+$ and $\SU^-$ will be regarded as light cone evolution operators.
Equation \rf{UfromQ} is easily proven by noting that
\begin{equation}
\SQ^+(s_+)\,=\,\SY_{\infty}^{-1}\cdot\prod_{n=1}^{\SRN}\,G_{2s}(\sf_{2n-1})\,,\qquad
(\SQ^-(s_-))^{-1}\,=\,\SY_{-\infty}^{}\cdot\prod_{n=1}^{\SRN}\,G_{2s}(\sf_{2n-1})\,.
\end{equation}
The operator $\SY_{\infty}$ satisfies
$\SY_{\infty}^{-1}\cdot\sf_{2n-1}^{}\cdot\SY_{\infty}^{}=\sf_{2n}$.
This implies
\[
\SQ^+(s_+)\cdot(\SQ^-(s_-))^{-1}\,=\,\prod_{n=1}^{\SRN}\,G_{2s}(\sf_{2n})\cdot\SY_{\infty}^{-1}\cdot\SY_{-\infty}\cdot\prod_{n=1}^{\SRN}\,G_{2s}(\sf_{2n-1})\,.
\]
It remains to notice that $\SY_{\infty}^{-1}\cdot\SY_{-\infty}=\SU_0$ to conclude the proof of \rf{UfromQ}.

\subsection{Chiral Q-operators in the  lattice KdV model}

Note that the Q-operators $\SQ^+_{\rm\sst KdV}$ and $\SQ^-_{\rm\sst KdV}$ are indeed the direct
massless limits of  $\SQ^+_{\rm\sst ShG}(s|u)\equiv\SQ^+_{\rm\sst ShG}(u)$ and $\SQ^-_{\rm\sst ShG}(s|u)\equiv\SQ^-_{\rm\sst ShG}(u)$, respectively,
\begin{equation}\label{limSGKdV}
\begin{aligned}
&\SQ^+_{\rm \sst KdV}(u)\,=\,\lim_{\de\ra\infty}\;\SQ^+_{\rm
\sst ShG}(s+\de|u+\de)\,,\\
&\SQ^-_{\rm\sst KdV}(u)\,=\,\lim_{\de\ra\infty}\;\SQ^-_{\rm \sst ShG}(s+\de|u-\de)\,.
\end{aligned}\end{equation}
The Baxter equations relate the Q-operators $\SQ^\ep$ with the T-operators $\ST^\ep$. In the case of KdV theory we had
seen that $\ST^+_{}$ and $\ST^-_{}$ depend only on right- and left-moving degrees of freedom $\sw_n^+$ and $\sw_n^-$,
respectively. This suggests that $\SQ^+_{}$ and $\SQ^-_{}$ should have the same property. And indeed,
it can be checked that
\begin{equation}
[\,\SQ^+_{}(u)\,,\,\sw_n^-\,]\,=\,0\,,\qquad[\,\SQ^-_{}(u)\,,\,\sw_n^+\,]\,=\,0\,,
\end{equation}
making clear that $\SQ^+_{}(u)$ and $\SQ^-_{}(u)$ depend on the
right- and left-moving degrees of freedom only.
This property implies in particular that
\begin{equation}
[\,\SQ^+_{}(u)\,,\,\spp_\0\,]\,=\,0\,,\qquad[\,\SQ^-_{}(u)\,,\,\spp_\0\,]\,=\,0\,,
\end{equation}
which means that $\SQ^+_{}(u)$ and $\SQ^-_{}(u)$ can be projected onto
$\CF_p^+$ and $\CF_p^-$, respectively. We will use the notation
$\SQ^+_p(u)$ and $\SQ^-_p(u)$ for the resulting operators
acting within $\CF_p^+$ and $\CF_p^-$, respectively.


\section{Analytic properties of Q-operators}\label{Qanal}

\setcounter{equation}{0}

It turns out
that the operators $\SQ^\ep(u)$ are hermitian up to a phase 
for $u\in\BR$, see Subsection \ref{herm} below for the precise
statement.
It follows that the T- and the Q-operators can be diagonalized simultaneously. To each eigenstate of the
evolution operator $\SU$, we therefore have a quadruple of functions $(t^+(u),q^+(u),t^-(u),q^-(u))$ related
to each other by equations of Baxter type, as written out explicitly in \rf{BAX} below.

Understanding the analytic properties of the Q-operators
or (equivalently) of their eigenvalues $q^\ep(u)$, $\ep=\pm$ 
is a key step towards understanding the
spectrum of the theories in question: It turns out that the analytic properties of the functions
$q^\ep(u)$ following from their explicit constructions restrict the relevant class of solutions
to the Baxter equations considerably. Let us call a pair of solutions
of the Baxter equations \rf{BAX} which has all these analytic properties {\it admissible}.
Being an admissible pair of solutions to the Baxter equations is clearly {\it necessary} for
functions $q^\ep(u)$, $\ep=\pm$ to represent eigenstates of $\SU$. The Separation of Variables Method
of Sklyanin, developed for the models of interest in the following section,
will then allow us to actually construct an eigenstate of $\SU$ to each pair
of admissible solutions to the Baxter equations. Being admissible is therefore
not only necessary, but also sufficient for solutions to the Baxter equations
$q^\ep$, $\ep=\pm$ to represent eigenstates of $\SU$.

\subsection{Analyticity}

The functions $q^\ep(u)$ are meromorphic with poles contained in the sets
\begin{equation}
\begin{aligned}
&\CS_{\ep s}\cup(-\CS_{\ep s}) & \;\;\text{for the Sinh-Gordon model,}\\
&\CS_{\ep s} & \;\;\text{for Liouville and KdV theory,}
\end{aligned}
\end{equation}
where the set $\CS_s$ is defined as
\begin{equation}
\CS_s\,=\,s-i\big(\eta+b\BZ^{\geq 0}+b^{-1}\BZ^{\geq 0}\big)\,.
\end{equation}
The proof is very similar to the one given in \cite[Section 4]{BT} for the case of the Sinh-Gordon model.

In the case of KdV theory we may furthermore discuss the 
dependence of the operators
$\SQ^\ep_p(u)$ with respect to the parameters $p$. 
It is meromorphic and analytic in the
strip
\begin{equation}
\BS_p\,=\,\{\,p\in\BC\,;\,|{\rm Im}(p)|< N\fr{Q}{2}\,\}\,.
\end{equation}
The proof becomes simple
if one uses the alternative integral operator representation
\rf{t-rep} for $\SQ^\ep_{\rm \sst KdV}(u)$ given in Appendix A.

\subsection{Asymptotics}

Probably the most important difference between the massive and the
massless cases concern the asymptotic properties of the Q-operators.
Whereas we can find exponential decay of the Q-operator at both ends of the
u-axis in the case of the Sinh-Gordon model,
\begin{equation}
q_{\rm \sst ShG}(u)\underset{\substack{|u|\ra\infty\\ 
{\rm Im}(u)={\rm const}}}{\sim}
e^{\pi i\SRN s|u|}e^{-{\pi}\SRN\eta |u|}\,,
\end{equation}
in the remaining cases we find exponential decay only at one and of the u-axis,
\begin{equation}\label{qasym1}
q^\ep(u)\underset{\substack{u\ra\ep\infty\\ 
{\rm Im}(u)={\rm const}}}{\sim}
e^{\pi i\SRN s|u|}e^{-{\pi}\SRN\eta|u|}\,,
\end{equation}
while we have oscillatory asymptotic behavior at the other end: There exists a real number $p$ and
constants $N^{\ep}$, $C^\ep(p)$ and $D^\ep(p)$ such that
 \begin{equation}\label{qasym2}
q^\ep(u)\underset{\substack{u\ra-\ep\infty\\ 
{\rm Im}(u)={\rm const}}}{\sim}
N^\ep\,e^{-\frac{\pi i}{2}\SRN u^2}\,\big(
C^\ep(p) \,e^{2\pi ip  u}+D^\ep(p)\,e^{-2\pi ip u}\big)\,.
\end{equation}

Most of the properties above can be proven by straightforward extensions of the
arguments in \cite{BT}.
This is not the case for the
oscillatory asymptotics \rf{qasym2}. We therefore give a sketch of the proof in Appendix A.

\subsection{Hermiticity}\label{herm}

Some of the 
properties of the Q-operators become most transparent in terms of the
modified Q-operators $\hat\SQ^\ep(u)$ which are defined as
\begin{equation}
\hat\SQ^\ep(u)\,=\,\Xi^\ep(u)\,\SQ^{\ep}(u)\,,
\end{equation}
with normalization factors $\Xi^\ep(u)$ 
being chosen as 
\begin{equation}
\begin{aligned}
& \Xi^{\ep}(u)\,=\,\left(\frac{F(u+\ep s-i\eta)}{F(u-\ep s+i\eta)}\right)^{\SRN}  \\
& \Xi^{\ep}(u)\,=\,\left(\frac{F(u+\ep s-i\eta)}{F_0(u-\ep s+i\eta)}\right)^{\SRN}
\end{aligned}
\quad\begin{aligned}
&\text{for the Sinh-Gordon model,}\\[2ex]
&\text{for Liouville and KdV theory.\,}
\end{aligned}
\end{equation}
with 
$F(v)=(F_0(v))^{-1}\Phi(v)$, where 
$F_0(x)=\zeta^2 e^{\frac{\pi i}{4}(x^2+\frac{1}{2})}$,
$\zeta=e^{\frac{\pi i}{24}(b^2+b^{-2})}$ and
\begin{align}
&
\Phi(x)\,=\,
\exp\left(\int_{\BR+i0}\frac{dt}{8t}\;\frac{e^{-2itx}}{\sinh (bt)\sinh (b ^{-1}t)\cosh ((b+b^{-1})t)}\right)\,.
\end{align}
The function $\Phi(x)$ was introduced in \cite{BMS}
\footnote{A relative had previously appeared in \cite{LZ}}, where
all properties relevant for us are listed in the appendix.

We then find that
the operators $\hat\SQ(u)$ are hermitian for all $u\in\BR$,
\begin{equation}
\big(\hat\SQ(u)\big)^{\dagger}\,=\,\hat\SQ(u)\qquad\forall u\in\BR\,.
\end{equation}
This can be verified by using the 
integral identity (A.31) in \cite{BT}, taking into account the
functional relation $F(x+i\eta)F(x-i\eta)=(w_b(x))^{-1}$ \cite{BMS}.

This property  implies in  particular that the coefficients $C^\ep(p)$ and $D^\ep(p)$ that appear in \rf{qasym2}
are complex conjugate to each other, $(C^\ep(p))^{\ast}=D^\ep(p)$.
Of particular interest will be the so-called
{\it reflection amplitude} defined by
\begin{equation}\label{Refdef}
R^\ep(p)\;=\;(C^\ep(p))^\ast\,/\,C^\ep(p)\,.
\end{equation}
This quantity will play an important role later.


\subsection{Functional relations}

\subsubsection{Baxter equations}

The $Q$-operators all satisfy Baxter-type finite difference equations of the general
form
\begin{equation}\label{BAX}
\ST(u)\SQ(u)\,=\,A(u)\SQ(u-ib)+D(u)\SQ(u+ib)\,.
\end{equation}
The coefficient functions $A(u)$ and $D(u)$ are model-dependent.
In the massive case (Sinh-Gordon model) we find
\begin{equation}
\begin{aligned}
&A^+(u)\,=\,A^-(u)\,=\,e^{-\pi b\SRN (u-\frac{i}{2}b)}
\big(1+e^{-2\pi b(s-u+\frac{i}{2}b)}\big)^\SRN\,,\\
&D^+(u)\,=\,D^-(u)\,=\,e^{+\pi b\SRN(u+\frac{i}{2}b)}
\big(1+e^{-2\pi b(s+u+\frac{i}{2}b)}\big)^\SRN\,,
\end{aligned}\end{equation}
whereas we have for the massless cases (Liouville theory, KdV model) the expressions
\begin{align}
\label{ADmassless}
& \begin{aligned}
&A^{+}(u)= 
e^{-\pi b\SRN(u-\frac{i}{2}b)}
\big(1+e^{-2\pi b(s-u+\frac{i}{2}b)}\big)^{\SRN}\\[.5ex]
&A^{-}(u)= e^{-\pi b\SRN(u-\frac{i}{2}b)}
\end{aligned}\!\!\!\!\!\!\!\!\!\!\!\!\!\!\!\!
\begin{aligned}
D^{+}(u)= e^{\pi b\SRN(u+\frac{i}{2}b)}\,, & \\[.5ex]
D^{-}(u)= 
e^{\pi b\SRN (u+\frac{i}{2}b)}\big(1+e^{-2\pi b(s+u+\frac{i}{2}b)}\big)^{\SRN}\,.&
\end{aligned}
\end{align}
The proof of the Baxter equations given in \cite{BT} 
for the case of the Sinh-Gordon model which is similar to the methods of
\cite{Ba73,BS,PG,De99}
can easily be extended to the other cases.

\subsubsection{Quantum Wronskian relations}

The following bilinear functional relation is particularly useful:
\begin{align}\label{q-Wr}
\hat\SQ(v+i\de_+) \,
\hat\SQ(v-i\de_+) -\hat\SQ(v+i\de_-) \, \hat\SQ(v-i\de_-)
\,=\,1\,.
\end{align}
This relation is often called the quantum Wronskian relation.
The proof of \rf{q-Wr} in the case of the Sinh-Gordon model \cite{BT}
can easily be extended to the other cases.

It is worth noting that the
quantum Wronskian relation fixes
the absolute value of the coefficient $C^\ep(p)$ which
appears in \rf{qasym2} to be
\begin{equation}\label{specmeas}
|C^\ep(p)|^2=({4\sinh (2\pi b p)\sinh (2\pi b^{-1} p)})^{-1}.
\end{equation}
The quantity $|C^{\ep}(p)|^{-2}$ will later be identified as a natural spectral measure.

\subsection{Scale invariance}

It is worth observing that
the dependence of $\SQ_{\rm \sst Liou}^{\ep}(s|u)\equiv\SQ_{\rm \sst Liou}^\ep(u)$, $\ep=\pm$ w.r.t. the
scale parameter $s$ can
(up to unitary equivalence) be absorbed into a shift of $u$,
\begin{equation}\label{scaleinv}
\begin{aligned}
&\SQ_{\rm\sst Liou}^{+}(s|u)\,=\,\SG^{-s}\cdot\SQ_{\rm \sst Liou}^{+}(0|u- s)\cdot \SG^{+s}\,,\\
&\SQ_{\rm\sst Liou}^{-}(s|u)\,=\,\SG^{-s}\cdot\SQ_{\rm \sst Liou}^{-}(0|u+ s)\cdot \SG^{+s}\,,
\end{aligned}\end{equation}
where $\SG$ is the unitary operator $\SG=\prod_{r=1}^{\SRN}\su_r^{-\frac{i}{b}}$.
A similar (even simpler) property
holds for $\SQ^{\ep}_{\rm\sst KdV}(u)$. This reflects the scale invariance of these theories.

Equation \rf{scaleinv} implies in particular that in the massless cases one may
represent the eigenvalues of $\SQ^+(u)$ and $\SQ^-(u)$ by functions
$q^+(u-s)$ and $q^-(u+s)$ which do not carry any dependence on $s$ other
than the one implied by the form of the arguments, respectively.

\section{Separation of variables}

\setcounter{equation}{0}

The construction of the Q-operator allowed us to deduce a set of
conditions that are necessary for functions $q^\ep(u)$ to represent
an eigenvalue of $\SQ^\ep(u)$. It remains to show that these
conditions are also sufficient, i.e. that to each solution
of these conditions there exists an eigenvector $\Psi_q\in\CH$
such that $\SQ^\ep(u)\Psi_q=q^\ep(u)\Psi_q$. We will now show how to
construct such an eigenvector with the help of the
separation of variables method \cite{Sk1,Sk2,Sk3}.
The upshot is to show
existence of a representation $\CH_{\rm \sst SOV}$ for $\CH$ in
which states $\Psi$ are
represented by wave-functions $\Psi(\by)$, $\by=(y_1,\dots,y_\SRN)$
such that eigenstates of the $\SQ^{\ep}(u)$ can be represented in a fully
factorized form
\begin{equation}
\Psi(\by)\,=\,\prod_{k=1}^{\SRN}q^{\ep(k)}(y_k)\,,
\end{equation}
for a certain choice of $\ep(k)$.
The wave functions $\Psi(\by)$ have to be normalizable w.r.t. to the
measure $d\mu(\by)$ which represents the scalar product in $\CH_{\rm\sst SOV}$. The
main issue is to show that the conditions on  $q^\ep(u)$ found
above ensure the normalizability w.r.t. $d\mu(\by)$.

In the case of the Sinh-Gordon model \cite{BT}
the representation $\CH_{\rm \sst SOV}$ is simply the spectral representation
for the commutative family of operators $\SB(u)$ defined as the off-diagonal
element of the monodromy matrix $M(u)=\big(\begin{smallmatrix}\SA(u) & \SB(u) \\
\SC(u) &\SD(u)
\end{smallmatrix}
\big)$.
We will now briefly discuss how to adapt this method to the
remaining cases.

\subsection{Separation of variables for the Liouville and quantum KdV theories}


The elements of the monodromy matrices $M^\ep(u)$, $\ep=\pm$, satisfy
the relations
\begin{align}\label{YBE}
R_{12}^{}(u-v)\,M^{\ep}_{1}(u)\, M^{\ep}_{2}(v)\,&=\,M^{\ep}_{2}(v)\,M^{\ep}_{1}(u)\,R_{12}^{}(u-v)\,,\\
R_{12}'(u-v)\,M_{1}^{+}(u)\, M_{2}^{-}(v)\,&=\,M^{-}_{2}(v)\,M^{+}_{1}(u)\,R_{12}'(u-v)\,,
\end{align}
where $R_{12}'(u)={\rm diag}(q,1,1,q)$ for KdV theory, while for Liouville 
theory
\begin{equation}
R_{12}'(u)=\left(
\begin{matrix} e^{\pi b(u+ib)} & & & \\
& e^{\pi b u} & 0 & \\
& i\sin\pi b^2 & e^{\pi b u} & \\
& & & e^{\pi b(u+ib)}
\end{matrix}\right),
\end{equation}
Let us use the notation $M^\ep(u)=\big(\begin{smallmatrix} \SA^\ep(u) & \SB^\ep(u)\\ \SC^\ep(u) &\SD^\ep(u)
\end{smallmatrix}\big)$.
The relations \rf{YBE} imply in particular that
\begin{equation}\begin{aligned}
\SB^\ep_{}(u)\SB^{\ep'}_{}(v)\,=\,\SB^{\ep'}_{}(v)\SB^{\ep}_{}(u)\,,\\
\SC^\ep_{}(u)\SC^{\ep'}_{}(v)\,=\,\SC^{\ep'}_{}(v)\SC^{\ep}_{}(u)\,,
\end{aligned}
\qquad\ep,\ep'=\pm\,.
\end{equation}
Note furthermore that $\SB^\ep(u)$, $\SC^{\ep'}(u)$ are positive self-adjoint
for all $u\in\BR+i/2b$.
We may therefore simultaneously diagonalize either one of the the commutative families of operators
$\SB^{\ep}(u)$, $\ep=\pm$  or  $\SC^\ep(u)$, $\ep=\pm$. The main idea of the Separation of
Variables method is to work within the spectral representation for one of these families.

Let us consider the spectral representation for the operators $\SB^{\ep}(u)$, $\ep=\pm$.
It will be called the B-representation.
One may parameterize
the corresponding eigenvalues as
\begin{equation}\label{b-y}
\begin{aligned}
&b^+(u)\,=\,-ie^{\pi b u}b_0\prod_{a=1}^{L}\big(1-e^{+2\pi b(u-y_a^+)}\big)\,,\\
&b^-(u)\,=\,-ie^{\pi b u}b_0\prod_{a=0}^{L}\big(1-e^{-2\pi b(u-y_a^-)}\big)\,,
\end{aligned}
\qquad b_0=\prod_{a=1}^{L}e^{\pi by_a^+}\prod_{a=0}^{L}e^{-\pi b y_a^-}\,.
\end{equation}
The spectral representation
for the operators $\SB^\ep(u)$, $\ep=\pm$ 
is therefore equivalent to
a representation in terms of wave-functions $\Psi(\by)$, 
where $\by=(\,y_{1}^+,\dots,y_L^+\,;\,y_0^-,y_1^-,\dots,y_L^-)$. Let us 
define operators $\sy_a^\ep$ such that
$\sy_a^\ep\cdot\Psi(\by)=y_a^\ep\Psi(\by)$.

Considering the operators $\SC^\ep(u)$, $\ep=\pm$ instead yields what will be called
the C-representation in terms
of variables $\tilde\by=(\,\tilde y_{1}^-,\dots,\tilde y_L^-\,;\,\tilde y_0^+,\tilde y_1^+,\dots,\tilde y_L^+)$.

\subsection{The Baxter equations}

\subsubsection{Liouville theory}

Let us define operators $\SA^\ep(\sy_a^{\ep})$, $\SD^\ep(\sy_a^{\ep})$ by the prescription to
order the operators $\sy_a^{\ep}$ to the left of the operators which appear in the expansion of
$\SA^\ep(u)$ in powers of $e^{\pi bu}$. It is an easy consequence of the algebraic relations
\rf{YBE} that these operators act on wave-functions $\Psi(\by)$ as finite difference operators
of the form
\begin{equation}\label{ADrepr}
\SA^\ep(\sy_a^{\ep})\cdot\Psi(\by)\,=\,A^\ep(y_a^{\ep})\,\de_{a-}^\ep\Psi(\by)\,,
\qquad
\SD^\ep(\sy_a^{\ep})\cdot\Psi(\by)\,=\,D^{\ep}(y_a^{\ep})\,
\de_{a+}^\ep\Psi(\by)\,,
\end{equation}
where $\de_{a\pm}^\ep$ are defined as
\[
\de_{a\pm}^{\ep}
\Psi(\dots,y_a^\ep,\dots)\,=\,\Psi(\dots,y_a^\ep\pm ib,\dots)\,.
\]
The coefficients $A^\ep(u)$, $D^\ep(u)$ are constrained by the quantum determinant condition
\begin{equation}
\Delta^\ep(u)\,\equiv\,\SA^{\ep}(u)\SD^\ep(u-ib)-\SB^\ep(u)\SC^\ep(u-ib)\,=\,\big(1+e^{-2\pi b(s-\ep(u-\frac{i}{2}b))}\big)^\SRN\,.
\end{equation}
As anticipated by the notation we shall adopt the 
choice \rf{ADmassless} for the
coefficients $A^\ep(u)$, $D^\ep(u)$.

The condition that $\Psi(\by)$ represents an eigenstate of the  transfer matrices $\ST^\ep(u)$, $\ep=\pm$, with eigenvalues
$t^{\ep}(u)$ becomes equivalent to the equations
\begin{align}
t^\ep(y_{a}^\ep)\,\Psi(\by)\,&=\,A^\ep(y_a^\ep)\,\de^\ep_{a-}
\Psi(\by)+D^\ep(y_a^\ep)\,\de_{a+}^\ep\Psi(\by)\,,\quad\ep=\pm\,.\label{BaxPsi}
\end{align}
The eigenfunctions for $\ST^{\ep}_{}(u)$ can therefore be constructed in the following form
\begin{equation}\label{factoransatz}
\Psi_{q}({\mathbf y})\,=\,\prod_{a=1}^{L}q^-_{}(y_{a}^+)\,\prod_{a=0}^{L}q^+_{}(y_{a}^-)\,,
\end{equation}
where $q_p^\ep(u)$, $\ep=\pm$ are solutions to the Baxter equations
\begin{equation}\label{pBax}
t^{\ep}_{}(u)\,q^\ep_{}(u)\,=\,A^\ep_{}(u)q^\ep_{}(u-ib)+D^\ep_{}(u)q^\ep_{}(u+ib)\,.
\end{equation}
Classifying eigenstates of $\ST^\ep(u)$, $\ep=\pm$ thereby becomes equivalent to finding the
proper set of solutions of the Baxter equations \rf{pBax}.

\subsubsection{KdV theory}

It is instructive to notice that the limit $s\ra\infty$ which yields the lattice KdV model from
Liouville theory forces one of the variables $y_a^-$, by convention
chosen to be the  variable $y_0^-\equiv y_\0$, to diverge.
The resulting parametrization for the eigenvalue $b^-(u)$ is
\begin{equation}
\begin{aligned}
&b^+(u)\,=\,-ie^{\pi b u}b_\0 e^{-\pi b y_\0} \prod_{a=1}^{L}\big(1-e^{+2\pi b(u-y_a^+)}\big)\,,\\
&b^-(u)\,=\,-ie^{\pi b u}b_\0 e^{+\pi b y_\0} \prod_{a=1}^{L}\big(1-e^{-2\pi b(u-y_a^-)}\big)\,,
\end{aligned}
\qquad b_\0=\prod_{a=1}^{L}e^{\pi by_a^+}\prod_{a=1}^{L}e^{-\pi b y_a^-}\,.
\end{equation}
The equations \rf{ADrepr} degenerate for $a=\0$ into 
\[
\SA^{\ep}(\sy_\0)\Psi(\by)\,=\,A^\0(\sy_\0)
\de_{\0-}\Psi(\by)\,,\qquad
\SD^{\ep}(\sy_\0)\Psi(\by)\,=\,D^\0(\sy_\0)
\de_{\0+}\Psi(\by)\,,
\]
where $A^\0(u)=e^{-\pi b \SRN(u-\frac{i}{2}b)}$,  
$D^\0(u)=e^{+\pi b \SRN(u+\frac{i}{2}b)}$, 
respectively, so that \rf{BaxPsi} for $a=\0$
becomes 
\begin{align}
t_\0^{}\,\Psi(\by)\,&=\,A^\0(y_\0)\de_{\0-}\Psi(\by)+D^\0(y_\0)
\de_{\0+}\Psi(\by)\,,
\label{BaxPsi0}
\end{align}
where $
t_\0^{}=t^+(-\infty)=t^-(\infty)$.
We accordingly need to modify \rf{factoransatz} as
\begin{equation}\label{factoransatz0}
\Psi_{q}({\mathbf y})\,=\,\prod_{a=1}^{L}q^-(y_{a}^+)\,q^\0_{}(y_\0)\prod_{a=1}^{L}q^+(y_{a}^-)\,.
\end{equation}
The equation \rf{BaxPsi0} is solved by the exponential functions 
$q^\0_{}(y_\0)=e^{-\frac{\pi i}{2}\SRN u^2}e^{2\pi i y_\0^{}p}$, with
$p$ being related to $t_\0$ as $t_\0=2\cosh(2\pi b p)$.
We will see that $p$ can take arbitrary real values.

\subsection{The Sklyanin measure}

Adopting the parametrization \rf{b-y} for the eigenvalues of the operators $\SB^\ep(u)$, $\ep=\pm$
one needs to find the set of all $\by\in\BC^{\SRN}$ which parameterize a point in the spectrum
of $\SB^\ep(u)$ via \rf{b-y}. We shall adopt the following conjecture:

\begin{conj}
All points in the spectrum of $\SB^\ep(u)$, $\ep=\pm$ can be parameterized by real values of $y_1^+,\dots,y_L^+$
and $y_0^-,\dots,y_L^-$.
\end{conj}

Validity of the conjecture above is {\it not} crucial for the discussion below, we adopt it here
to simplify the exposition. However, we are rather confident that it is correct. It
can be checked in certain limits and special cases.
The conjecture implies that the
B-representation can be realized on a Hilbert space of the form
\[
\CH_{\rm\sst SoV}^{\SB}\,=\,L^2\big((\BR^{L}/S_L)\times(\BR^{L+1}/S_{L+1})\,;\,d\mu_\SB^{}\big)\,.
\]
Elements of $\CH_{\rm\sst SoV}^\SB$ are represented by wave-functions
$\Psi(\by)$ that are normalizable w.r.t. $d\mu_\SB$ and totally symmetric
under permutations among the sets of variables $\{y_a^+;a=1,\dots L\}$ and
$\{y_a^-;a=0,\dots L\}$, respectively.
The C-representation can similarly be realized on
\[
\CH_{\rm\sst SOV}^{\SC}\,=\,
L^2\big((\BR^{L+1}/S_{L+1})\times(\BR^{L}/S_{L})\,;\,d\mu_\SC^{}\big)\,,
\]
Elements of $\CH_{\rm\sst SoV}^\SC$ are represented by wave-functions
$\Psi(\tilde\by)$ that are normalizable w.r.t. $d\mu_\SC$ and totally symmetric
under permutations among the sets of variables $\{\tilde y_a^+;a=0,\dots L\}$ and
$\{\tilde y_a^-;a=1,\dots L\}$, respectively.

The Sklyanin measure $d\mu_\SB$ 
can be found by the same method as used in \cite{BT} 
from the requirement that $\SA^\ep(v)$ and
$\SD^\ep(v)$ are positive self-adjoint. We have
\begin{equation}\label{Sklmeas}
d\mu_{\SB}^{}(\by)\,=\,d\mu_{\SB}^+(\by^+)\,d\mu_{\SB}^-(\by^-)\,,
\end{equation}
where
\begin{equation*}
\begin{aligned}
& L!\, d\mu_{\SB}^+(\by^+)\,=\,
\prod_{a=1}^{L}dy_a^{+}\; e^{\pi Q (L+1) y_a^+}\,
\prod_{b<a} 2\sinh\pi b(y_{a}^++y_{b}^+)
2\sinh\pi b^{-1}(y_{a}^{+}-y_{b}^{+})\,,\\
& (L+1)!\,d\mu_{\SB}^-(\by^-)\,=\,
\prod_{a=0}^{L}dy_a^{-}\;e^{\pi QLy_a^-}\prod_{b<a}
2\sinh\pi b(y_{a}^--y_{b}^-)2\sinh\pi b^{-1}(y_{a}^{-}-y_{b}^{-})
\,.\end{aligned}
\end{equation*}
We have a very similar expression for $d\mu_{\SC}^{}(\by)$.

In the case of the lattice KdV theory we get the following modifications:
\begin{equation}\label{SklmeasKdV}
d\mu_{\SB}^{}(\by)\,=\,d\mu_{\SB}^+(\by^+)\,dy_\0\,d\mu_{\SB}^-(\by^-)\,,
\end{equation}
where $d\mu_{\SB}^+(\by^+)$ is unchanged, but $d\mu_{\SB}^-(\by^-)$
is now given as
\begin{equation*}
L!\,d\mu_{\SB}^-(\by^-)\,=\,
\prod_{a=1}^{L}dy_a^{-}\;e^{\pi Q(L+1)y_a^-}\prod_{b<a}2
\sinh\pi b(y_{a}^--y_{b}^-)2\sinh\pi b^{-1}(y_{a}^{-}-y_{b}^{-})\,.
\end{equation*}
It is worth observing that the small asymmetry between the 
Liouville-variables
$y_a^+$ and $y_a^-$ 
disappears in the limit giving quantum KdV theory. 

\section{The spectra}

\setcounter{equation}{0}

\subsection{The spectrum of quantum KdV theory}

The fact that the dynamics generated by $\SU_{\rm\sst KdV}$ is "trivial" in the sense
that it decouples into right- and left motions \rf{w-dyn} of $\sw_{\nu,t}^+$ and $\sw_{\nu,t}^-$ respectively,
does not mean that the lattice model
characterized by the T-operators $\ST_{\rm\sst KdV}^\ep$, $\ep=\pm$,
is trivial as an integrable model. As in classical
(m)KdV theory one may define alternative and much less trivial evolutions from the families of operators $\ST_{\rm\sst KdV}^\ep$
or $\SQ_{\rm\sst KdV}^\ep$.
The diagonalization of these operators is interesting
in its own right.

\subsubsection{The spectrum of the chiral free field}

Let us first study the chiral free field theories with Hilbert space $\CF_p^\ep$ and Q-operator $\SQ^\ep_p(u)$ for
fixed values of $\ep\in\{\pm\}$ and $p\in\BR$.
The spectral theorem for the commutative family of self-adjoint operators
$\hat\SQ^{\ep}_p(u)$ implies that 
the eigenstates $f_q^\ep\in\CF_p^\ep$ of these operators
form a {\it basis} for $\CF_p^\ep$.
This is the case for arbitrary real values of the variable $p$.
Let $q_{p}^{\ep}(u)$ be the eigenvalue of the operator $\SQ^{\ep}_p(u)$ on 
$f_q^\ep$.
It must be element of the set
$\CQ_{p}^{\ep}$, the  set of all functions $q_{p}^{\ep}(u)$ that possess all the analytic and
asymptotic properties implied by our explicit construction of the Q-operators as discussed in
Section \ref{Qanal}.

On the other hand let let us note that
the SOV representation is realized on the Hilbert spaces
\begin{equation}
\CH_{\rm\sst SoV}^\ep\,=\,L^2(\BR^{L};d\mu_{\SB}^{\ep})_{\rm\sst symm}^{}\,.
\end{equation}
For a given element $q_{p}^{\ep}(u)\in \CQ_{p}^{\ep}$ define
\begin{equation}\label{SOVpm}
\Psi_{q}^{\ep}(\by^\ep)\,=\,\prod_{a=1}^{L}q_{p}^{\ep}(y^{\ep}_a)\,.
\end{equation}
It follows from the asymptotic properties of $q_{p}^{\ep}(u)$ that $\Psi_{q}^{\ep}(\by^\ep)$ is normalizable w.r.t. $d\mu_\SB^\ep$. 
There is a corresponding eigenstate $f_q^\ep\in\CF_p^\ep$ of
$\SQ^\ep_p(u)$ which has as its eigenvalue the function $q_{p}^{\ep}(u)$
we had used in \rf{SOVpm}. We conclude that there is a one-to-one correspondence between the
elements of $\CQ_{p}^{\ep}$ and the eigenstates of $\SQ^{\ep}_p(u)$ within $\CF_p^\ep$.
The fact that the wave-function $\Psi_{q}^{\ep}$ are all 
normalizable implies in particular
that the spectrum of $\SQ^{\ep}_p(u)$ is {\it purely discrete}.

\subsubsection{The zero mode spectrum of quantum KdV theory}

To each triple
$q\equiv(q_{p}^{+}(u),q_{p}^{\0}(u),q_{p}^{-}(v))$
of solutions to the Baxter equations \rf{BaxPsi}
we may associate a wave-function of the form
\begin{equation}\label{factoransatz-K}
\Psi_{q_p}({\mathbf y})\,=\,\prod_{a=1}^{L}q_{p}^-(y_{a}^-)\,q_p^\0(y_0^{})\,\prod_{a=1}^{L}q_{p}^+(y_{a}^+)\,.
\end{equation}
The asymptotic behavior \rf{qasym1}, \rf{qasym2} ensures the (plane-wave) normalizability
of $\Psi_q(\by)$.
We need to identify the set of solutions
of the zero  mode equation \rf{BaxPsi0} which yields a {\it complete}
set of $\SQ^\ep_{\rm\sst KdV}$-eigenstates in this way.

By means of induction it is easy to prove that $\ST_0^{{}}$ has the following form:
\begin{align}
&\ST_0\,=\,2\cosh\pi b \spp_\0\,,\label{ST0forma}
\end{align}
where $e^{2\pi b \spp_\0}\equiv\prod_{n=1}^{\SRN}\su_n$. It easily follows from this observation
that the vectors $\Psi_q(\by)$ constructed from the
choices $q_{p}^{\0}(u)=e^{-\frac{\pi i}{2}u^2}e^{2\pi ipu}$, $p\in\BR$,
all represent linearly independent basis vectors for  $\CH$ in the sense of generalized functions.

\subsection{The spectrum of Liouville theory}

We are now going to analyze the spectrum of Liouville theory in a similar manner.
To each eigenstate $\Psi$
of the Q-operators $\SQ_{}^+(u)$ and  $\SQ_{}^-(u)$ there exists a
complex number $p$ and a corresponding pair of elements
$q_p=(q^+_p,q^-_p)\in\CQ_p^+\times\CQ_p^-$, given by the eigenvalues of $\SQ_{}^\ep(u)$ on $\Psi$.
Conversely, for a given value of $p$ and each pair $q_p=(q_p^+,q_p^-)\in\CQ_p^+\times\CQ_p^-$ of admissible solutions to the Baxter equations
one may construct an eigenstate of
the Q-operators $\SQ_{}^+(u)$ and  $\SQ_{}^-(u)$ as
\begin{equation}\label{factoransatz-L}
\Psi_{q_p}({\mathbf y})\,=\,\prod_{a=0}^{L}q_{p}^-(y_{a}^-)\,\prod_{b=1}^{L}q_{p}^+(y_{b}^+)\,.
\end{equation}
With the help of our explicit formulae for the Sklyanin measure and the
formulae \rf{qasym1}, \rf{qasym2} for the asymptotic behavior of the functions $q_p^{\ep}(u)$
it is possible to check that the states \rf{factoransatz-L} are plane-wave normalizable if $p\in\BR$.
More precisely one may show that 
\begin{equation}
\big(\,\Psi_{q_p}\,,\,\Psi_{q_{p'}}\,\big)\,=\,\frac{\de(p-p')}{4\sinh (2\pi b p)\sinh (2\pi b^{-1} p)}\,.
\end{equation}
This means that $dp \,4\sinh (2\pi b p)\sinh (2\pi b^{-1} p)$ is the natural spectral measure 
for the integration over $p$ in  the spectral representation.

One should note
that the spectrum of the zero mode $p$ is real and {\it purely continuous}.
This follows from the works \cite{Ka00,FK}, one of the main results of which
can be stated as
\begin{equation}
{\rm Spec}(\SU^+)\,=\,\big\{\,e^{-2\pi i(\De_p+m)/\SRN}\,;\,p\in\BR_+\,,
\,m\in\BZ/\SRN\BZ\,\big\}\,,
\end{equation}
where
\begin{equation}
\De_s\,=\,\frac{c-1}{24}+s^2\,,\qquad c=1+24\eta^2\,.
\end{equation}

It is
an important difference to the case of KdV theory that the eigenstates $\Psi_{q_p}$ and $\Psi_{q_{-p}}$ are not independent.
Indeed, it follows easily from \rf{t-rep} that the $q_{p}^{\ep}(u)$ are symmetric w.r.t.
$p$, i.e. $q_{p}^{\ep}(u)=q_{-p}^{\ep}(u)$. It follows that
\begin{equation}\label{refprop}
\Psi_{q_p}({\mathbf y})\,=\,\Psi_{q_{-p}}({\mathbf y})\,.
\end{equation}
We conclude that there is a one-to-one correspondence between triples $q=(p,q_p^+,q_p^-)$, $p\in\BR^+$,
$(q_p^+,q_p^-)\in\CQ_p^+\times\CQ_p^-$ and the elements of a basis for $\CH$ consisting
of generalized eigenstates of the Q-operators.

\section{The relation between quantum Liouville- and KdV-theory}

\subsection{The B\"acklund transformations}\label{Baeck}

The key point for us to observe is the fact that the sets $\CQ_p^\ep$ of admissible solutions of the
Baxter equations are the same for Liouville theory and the quantum lattice KdV model.
We may therefore construct operators 
$\SW_\chi$ which send the eigenstate 
$\Psi_{q}$ of 
$\SQ_{\rm\sst Liou}^\ep(u)$, $\ep=\pm$
associated to a triple $q=(q;q_p^+,q_p^-)$
to the eigenstate $\Phi_{q}$ of  
$\SQ_{\rm\sst KdV}^\ep(u)$, $\ep=\pm$, which in the
$\SB_{\rm\sst KdV}$-representation is represented 
by the wave-function
\begin{equation}\label{factoransatz-K'}
\Phi_{q}\,=\,
W_{\chi_q}\,
\prod_{a=1}^{L}q_{p}^-(y_{a}^-)\,q_p^\0(y_\0)
\prod_{b=1}^{L}q_{p}^+(y_{b}^+)\,,\quad
q_p^\0(y_\0)\,=\,e^{-\frac{\pi i}{2}\SRN y_\0^2}e^{2\pi p y_\0}\,.
\end{equation}
The prefactor $W_{\chi_q}$ is required to 
satisfy $|W_{\chi_q}|^2=4\sinh (2\pi b p)\sinh (2\pi b^{-1} p)$
while its phase $e^{2i\chi_q}\equiv W_{\chi_q}^{}/W_{\chi_q}^*$
is left arbitrary for the moment.
The operators $\SW_\chi$ clearly satisfy
\begin{equation}
\boxed{\qquad
\SW_\chi\cdot\SQ_{\rm\sst Liou}^\ep(u)\,=\,
\SQ_{\rm\sst KdV}^\ep(u)\cdot \SW_\chi\qquad}
\end{equation}
and they define unitary operators 
$\check{\SW}_\chi$ from $\CH$ to the subspace $\CH_+$
of $\CH$ on which the zero mode momentum $\spp_\0$ is positive. 
The operators $\SW_\chi$ 
can be seen as representatives for (generalizations of the) 
quantum B\"acklund transformations which map the 
interacting dynamics of Liouville theory to the free field
dynamics. They make the decoupling of left- and right-moving 
degrees of freedom in Liouville theory
manifest.

\subsection{Relation with scattering theory}

All what is nontrivial about Liouville theory is hidden in the way the decoupling between left-and right-movers is 
disguised when studying its dynamics in terms of the original degrees of freedom $\pi_n$, $\phi_n$. The operators
$\SW_{\chi}$ which trivialize the dynamics are 
rather nontrivial objects for which we do not have an explicit 
representation at the moment.\footnote{Finding a more explicit representation would become possible once
we had an explicit representation for the transformation from the original to the separated variables.}
In the following we shall propose an interpretation of 
one of these operators related to the asymptotic
behavior of the time evolution.

\subsubsection{Wave- and scattering operators}\label{Sdef}

One should note that the operators $\SQ^{\ep}_{\rm\sst Liou}(u)$
and $\SQ^{\ep}_{\rm\sst KdV}(u)$ coincide in the limit where the zero mode $\phi_\0$ tends to infinity,
\begin{equation}\label{Zeroasym}
\lim_{\rho\ra\infty}\,\bra\,\Psi_q\,,\,\SQ^\ep_{\rm\sst Liou}(u)\Phi_{\rho}\,\ket\,=\,
\lim_{\rho\ra\infty}\,\bra\,\Psi_q\,,\,\SQ^\ep_{\rm\sst KdV}(u)\Phi_{\rho}\,\ket\,,
\end{equation}
for any wave-packet $\Phi_\rho$ that has support localized around $\phi_\0=\rho$. 
We have, in particular, a similar statement for the evolution operator $\SU$. 
It then follows from standard arguments 
that wave-packets for time $\tau\ra\pm\infty$ are always pushed into
the asymptotic region $\phi_0\ra\infty$ where the dynamics becomes the free field dynamics.
We may therefore define natural analogs of the  wave operators from quantum mechanical
scattering theory as
\begin{equation}
\SW_{+\infty}\,=\,\lim_{\tau\ra\infty}\,
(\SU_{\rm\sst KdV})^{-\frac{\tau}{2}}\cdot(\SU_{\rm\sst Liou})^{+\frac{\tau}{2}}\,,\quad
\SW_{-\infty}\,=\,\lim_{\tau\ra\infty}\,(\SU_{\rm\sst KdV})^{+\frac{\tau}{2}}\cdot(\SU_{\rm\sst Liou})^{-\frac{\tau}{2}}\,.
\end{equation} 
The operators $\SW_{\pm\infty}$ are easily seen to represent 
a particular case of the B\"acklund transformations introduced in Subsection
\ref{Baeck} above.

The scattering operator $\SS$ which maps the asymptotic shape of a wave packet for 
$\tau\ra-\infty$ to the one for $\tau\ra\infty$ can then be
defined as $\SS\equiv \SW_{+\infty}^{}\cdot\SW_{-\infty}^{-1}$. It can be
described in terms of its eigenvalues $S_{q_p}$ in the spectral representation.

\subsubsection{Relation to space asymptotics of wave-functions}\label{Reflop}

In quantum mechanical scattering theory there exist well-known results relating the
scattering operator $\SS$ to the (target-) space asymptotics of eigenfunctions 
of the corresponding Hamiltonian. It seems fairly clear that similar relations 
will hold in the present context, as now to be formulated more explicitly. 
We'd like to  analyze the representation of eigenstates $\Psi_q$ in the
zero mode Schr\"odinger representation where they are
represented by wave-functions $\Psi_q(\phi_\0)$ taking values in $\CF^+_{\phi_\0}\ot\CF^-_{\phi_\0}$.
It follows from \rf{Zeroasym} that the asymptotic behavior for $\phi_\0\ra\infty$ of the wave-functions $\Psi_q(\phi_\0)$
can be expanded into the eigenstates of $\SQ^{\ep}_{\rm\sst KdV}(u)$,
\begin{equation}\label{Psiasym}
\Psi_{q_p}(\phi_\0)\,\underset{\phi_0\ra\infty}\sim\,N_p\,\big[\,e^{2\pi ip\phi_\0}+S_{q_p}e^{-2\pi ip\phi_\0}\,\big]\, (\,f_{q}^{+}\ot f_{q}^{-}\,)\,,
\end{equation}
where  $N_p$ is a normalization factor and $f_{q}^+\ot f_{q}^-\in\CF^+_p\ot\CF^-_p$
is an eigenstate of both $\SQ^+_{\rm\sst KdV}(u)$ and $\SQ^-_{\rm\sst KdV}(u)$ with eigenvalues
$q_{p}^+(u)$ and $q_{p}^-(u)$, respectively. We claim that the so-called reflection amplitudes $S_{q_p}$ which appear in the
asymptotic behavior \rf{Psiasym} are indeed the eigenvalues of the scattering operator $\SS$ defined
above.

\subsection{Relation between the reflection amplitudes of Liouville and of KdV theory}

Let us finally note that there is a remarkable relationship between the scattering
amplitude $S_{q_p}$
of Liouville theory and the reflection phases
$R^\ep(p)$ of KdV-theory introduced in \rf{Refdef},
\begin{equation}\label{Sfactor}
{\quad
S_{q_p^{}}\,\;=\;R_{q_p^+}\,R_{q_p^-} \quad{\rm if}\quad q_p=(q_p^+(u),q_p^-(u))\,.\quad}
\end{equation}
We have used the notation $R_{q_p^\ep}$, $\ep=\pm$ for the ratio
$R^\ep(p)=(C^\ep(p))^\ast/C^\ep(p)$
of the coefficients which appear in the asymptotic behavior of 
$q_p^{\ep}(u)$
for $u\ra -\ep\infty$ according to \rf{qasym2}.

The relationship \rf{Sfactor} allows one to calculate the scattering
operator $\SS$ from the asymptotics of the operators 
$\SQ^\ep_{\rm\sst KdV}(u)$ as determined in the Appendix. 
We do not go further into this direction for the case of the lattice 
models as we did not yet find a sufficiently nice formula for $\SS$. 
The situation becomes better in the continuum limit where \rf{Sfactor}
will be a key ingredient in our calculation of the Liouville reflection
amplitude. 

\subsubsection{Derivation of equation \rf{Sfactor}}

Equation \rf{Sfactor} can be verified by means of arguments which are similar to those in
\cite{T}. One may analyze the massless limit $s\ra\infty$ in two different ways.

Let us, on the one hand, consider an eigenstate $\Psi_q$ in the Sinh-Gordon model
represented in the Schr\"odinger representation
by a wave-function $\Psi_q(\phi_\0)\in\CF^+_{\phi_\0}\ot\CF^-_{\phi_\0}$.
Note that the limit giving Liouville theory from the Sinh-Gordon model
combines the limit $s\ra\infty$ with $\phi_\0\ra -\infty$.
It follows that the limit of the operator $\SQ_{\rm\sst ShG}(u)$ for $s\ra\infty$
can also be regarded as the
asymptotic behavior of $\SQ_{\rm\sst Liou}^{\ep}(u)$ for $\phi_\0\ra \infty$.
Arguing as in Subsection \ref{Reflop} we conclude  that
the leading behavior of $\Psi_q(\phi_\0)$ for $s\ra\infty$
can be described in terms of
eigenfunctions of $\SQ_{\rm\sst Liou}^\ep(u)$ as
\begin{equation}
\Psi_q(\phi_\0)\,\simeq\,
(\,C_{q_p}^{}\,e^{2\pi ip\phi_\0}+C_{q_p}^\ast\,e^{-2\pi ip\phi_\0}\,)
\, (\,f_{q}^{+}\ot f_{q}^{-}\,)\,,
\end{equation}
where  $f_{q}^+\ot f_{q}^-\in\CF^+_p\ot\CF^-_p$
is an eigenstate of both $\SQ^+_{\rm\sst KdV}(u)$ and $\SQ^-_{\rm\sst KdV}(u)$ with eigenvalues
$q_{p}^+(u)$ and $q_{p}^-(u)$, respectively. The eigenstate $\Psi_q$ is either
even or odd under parity. In order to evaluate this condition
note that $\arg S_{q_p}=-2\arg C_{q_p}=\rho_q(p)-4\pi ps$, where
$\rho_q(p)$ is independent of $s$.
For $s\ra\infty$ one gets the quantization condition
to leading order as the condition that there exists an integer $n$ such that
allowed values $p_n$ of the variable $p$ satisfy
\begin{equation}\label{quant1}
4\pi s\, p_n-\rho_q(p_n)\,=\,\pi n\,.
\end{equation}

One may, on the other hand, note that the limit $s\ra\infty$ of the 
Q-operators $\SQ_{\rm\sst ShG}^\ep(u)$ for
$s\ra\infty$
may according to \rf{limSGKdV}
be described either as the asymptotics of the $\SQ_{\rm\sst KdV}^+(u)$ for $u\ra -\infty$
or, equivalently as the asymptotics of $\SQ_{\rm\sst KdV}^-(u)$ for $u\ra +\infty$.
This implies for the eigenvalues of $\SQ_{\rm\sst ShG}^\ep(u)$
that we have, on the one hand
\begin{equation}
q^\ep(u)\,\simeq\,N_p\cos\big(2\pi p(u-s)+\theta_q^+(p)\big)\,,
\end{equation}
where $N_p=(\sinh (2\pi b p)\sinh(2\pi b^{-1}p))^{-\frac{1}{2}}$,
and on the other hand
\begin{equation}
q^\ep(u)\,\simeq\,N_p\cos\big(2\pi p(u+s)-\theta_q^-(p)\big)\,.
\end{equation}
The compatibility between these two equations requires that there exists an integer $n$ such that
\begin{equation}\label{quant2}
4\pi p_ns-\theta_q^+(p_n)-\theta_q^-(p_n)\,=\,\pi n\,.
\end{equation}
The equivalence of \rf{quant1} and \rf{quant2} 
yields our claim \rf{Sfactor}.

\subsubsection{Interpretation of equation \rf{Sfactor}}

It seems natural to interpret \rf{Sfactor} in the following way: 
In the same way as we used the 
evolution operator $\SU$ to define 
the scattering operator $\SS$ in Subsection \ref{Sdef} above, we may use the 
light-cone evolution operators $\SU^\ep$ to define light-cone scattering
operators $\SS^{\ep}$ for $\ep=\pm$, respectively. It is clear that the eigenvalues 
of the operators $\SS^+$ in a state defined by a pair $q_p=(q_p^+,q_p^-)$ will not depend on
$q_p^-$, and similarly for the eigenvalues of $\SS^-$.
It seems natural to conjecture that the eigenvalues of $\SS^{\ep}$ are precisely the phases
$R_{q_p^{\ep}}$ defined from the asymptotic behavior \rf{qasym2}
of $q_p^\ep$. This would mean that our relationship \rf{Sfactor} is equivalent to 
$\SS=\SS^+\SS^-$ which trivially follows from the factorization $\SU=\SU^+\SU^-$
observed in \rf{UfromQ} above.

\section{Continuum limit}

\setcounter{equation}{0}

Following arguments which are very similar to those used in \cite{T} we may now
reformulate the conditions for the q-functions in terms of nonlinear integral equations which generalize the equations coming from the
thermodynamic Bethe ansatz \cite{YY,Z1,Z2} to arbitrary excited states.
As shown in \cite{T},  one gets a characterization of the spectrum which is
{\it completely} equivalent to the one derived above. On the level of the
nonlinear integral equations it turns out to be straightforward to pass to the
continuum limit.
The limit is taken in such a way that $\SRN\ra\infty$, $s\ra\infty$ such that
\begin{equation}\label{contlim}
{mR}\,=\,4\,\sin\vt_0\,\SRN\,e^{-\pi b s}\,,\qquad\vt_0\equiv \frac{\pi b^2}{1+b^2}
\end{equation}
is kept constant.
As the necessary arguments are very similar to those in \cite{T}
we will only briefly
describe the resulting description of the q-functions for the
continuum theories and some of the most important consequences for the spectrum of
these theories.

\subsection{Reformulation in terms of integral equations}

As advertised earlier, one may express the eigenvalues of the
Q-operators in terms of the solutions of certain nonlinear
integral equations. 
These equations are
best formulated in terms of the functions
\begin{equation}
Y^\ep_p\big(\fr{\pi }{2\eta}u\big)\,=\,
q_p^\ep(u+i\de)q_p^\ep(u-i\de)\,,
\end{equation}
where $2\de=b^{-1}-b$. It suffices to consider the case that $p$ is purely imaginary
which is related the case of real $p$ by means of analytic continuation.
Assume that $q_p^\ep(u)$ has $M^\ep$ real zeros at positions $\vt_{a}^\ep$, $a=1,\dots,M$.
The functions
$q^\ep_p(u)$ can then be recovered from
\begin{equation}\label{qfromY}\begin{aligned}
\pa_\vt^{} \log  q^\ep_{p}\big(2\fr{\eta}{\pi} \vt\big)\,=\,
-\ep \,\frac{mRe^{\ep \vt}}{2\sin\vt_0}+ 
\sum_{a=1}^{M^\ep}&\;\frac{1}{\sinh(\vt-\vt^{\ep}_{a})}\\
& +\int_{\BR}
\frac{d\vartheta'}{4\pi}\;\frac{1}{\cosh(\vt-\vt')}\,\pa_{\vt'}^{}
\log\big(1+Y^\ep_p(\vt')\big)
\,,
\end{aligned}\end{equation}
The nonlinear integral equations in question
have an almost universal form,
\begin{equation}\label{ETBA}
\begin{aligned}
\log  Y^\ep_p(\vartheta) \,= -mRe^{\ep\vt}+\sum_{a=1}^{M^\ep} & \log S(\vartheta-\vartheta^{\ep}_{a}-i\fr{\pi}{2})\\[-1ex]
&\qquad
+\int_{\BR}
\frac{d\vartheta'}{4\pi}\; \si(\vartheta-\vartheta') \log(1+Y^\ep_p
(\vartheta'))\,,
\end{aligned}
\end{equation}
where
\[
 \si(\vt)\,=\,\frac{d}{d\vt}S(\vt)\,=\,\frac{4\sin\vt_0\cosh\vt}{\cosh 2\vt-\cos2\vt_0}\,.
\]
It is possible to prove that for arbitrary given input data $\bt^\ep=(\vt_{1}^\ep\dots,\vt_{M^\ep}^\ep)$, $\vt_{a}^\ep\in\BR$
the nonlinear integral equations
\rf{ETBA} have a unique solution $Y^{\ep}_{p,\bt}(\vt)$ which grows for $\vt\ra -\ep\infty$
as $2\pi\ep\,ip\,\vt$.\footnote{Bear in mind that we assume $p\in i\BR$.}
The equations \rf{ETBA} have to be supplemented by the set of equations
\begin{equation}\label{YBethe}
\begin{aligned}
2\pi\ep\, k_{a}^\ep\,=\,\ep \,mRe^{\ep\vt_a^\ep}
+\sum_{\substack{b=1\\b\neq a}}^{M^\ep}&
\arg S(\vartheta_a^\ep-\vartheta_b^\ep)\\[-2ex]
&\qquad +\int_{\BR}\frac{d\vartheta}{4\pi}\,\tau(\vartheta_a^\ep-\vartheta) \log(1+Y_{p,\bt}^\ep(\vartheta))
\,,
\end{aligned}
\end{equation}
where
\begin{equation}
\tau(\vartheta)\,\equiv\,
\frac{4\sin\vt_0\sinh\vt}{\cosh 2\vt+\cos2\vt_0}
=i\si(\vartheta+i\fr{\pi}{2})\,.
\end{equation}
The equations \rf{YBethe} represent strong constraints on the
parameters $\bt^\ep$. 
The fact that these parameters can only be real can be proven by means
of an argument similar to the one of \cite{YY,T} using the fact that the functions $Y^\ep_p(\vt)$
have to be real. This in turn follows from the hermiticity of the  Q-operators observed above.
In the following we shall adopt the basic conjecture that there exists a unique
solution to the equations \rf{YBethe} for any given tuples $\bk^{\ep}=(k^{\ep}_{1},\dots,k^{\ep}_{M^{\ep}})$.
If so, we can conclude that eigenstates are uniquely labelled
by $p$ and the tuples $\bk^\ep$.

\subsection{Analytic properties of the q-functions for the continuum theories}

The integral equations characterizing the q-functions
of the continuum theories are equivalent to either of the
following two functional equations,
\begin{align}
& t^\ep(u)q^\ep(u)\,=\,q^\ep(u+ib)+q^\ep(u-ib)\,,\label{CBax}\\
& q^\ep(v+i\eta)q^\ep(v-i\eta)-
q^\ep(v+i\de)q^\ep(v-i\de)\,=\, 1\,.\label{CWron}
\end{align}
We observe no difference between the massive and the massless cases.

The analytic properties of the q-functions also simplify in the continuum
limit. We find:
\begin{equation}\label{Canal}
\begin{aligned}
{\rm (i)} & \;\;\text{The q-functions are entire analytic in $u$ for
each of the cases considered.}\\
{\rm (ii)} & \;\;\text{The q-functions $q_p^{\ep}(u)$ 
are entire analytic in $p$ for
Liouville and KdV theory.}
\end{aligned}
\end{equation}
Important differences appear on the level of the asymptotic properties,
as we shall now discuss.
In the massive case we find \cite{T} rapid decay of $q^\ep(u)$ at both ends
of the real axis, more precisely,
\begin{equation}
\log q^\ep(u)\underset{{\rm Re}(u)\ra\pm \infty}{\sim}\,
-\frac{mR}{2\sin\vt_0}e^{\frac{\pi}{2\eta}|u|}\quad{\rm for}\;\;
|{\rm Im}(u)|<\eta\,.
\end{equation}
The decay of $q^\ep(u)$ implies that the spectrum of the Sinh-Gordon
field theory is
purely discrete.

As in the case of the lattice theory, the main difference to the
massless case is the appearance of oscillatory asymptotics at one
end of the real axis, while it remains rapidly decaying at the other end,
\begin{equation}\label{Casym}
\begin{aligned}
 q^\ep_p(u)\underset{{\rm Re}(u)\ra -\ep\infty}{\sim}\,
 &\frac{\cos(2\pi p u+\ep\theta_q(p))}
{\sqrt{\sinh(2\pi bp)\sinh(2\pi b^{-1}p)}} \\
 \log q^\ep_p(u)\underset{{\rm Re}(u)\ra\ep\infty}{\sim}\,
& -\frac{mR}{2\sin\vt_0}e^{\frac{\pi}{2\eta}|u|}
\end{aligned}
\quad{\rm for}\;\;
|{\rm Im}(u)|<\eta\,.
\end{equation}
One may formulate the above statements about the asymptotics of the
q-functions $q^\ep(u)$ for $u\ra \ep\infty$
more precisely by saying that there exists an asymptotic expansion of the
form
\begin{equation}
\log q^\ep_p(u)\,\sim\,-c_0\,e^{\frac{\pi}{2\eta}|u|}-
\sum_{n=1}^{\infty}c_n\,\BI_n^{\ep}\,e^{-\frac{\pi}{2\eta}(2n-1)|u|}\,.
\end{equation}
For the classical continuum field theories it is well-known that the
coefficients $\BI_n^\ep$ represent the {\it local} conserved quantitites of the
model in question. The coefficients $\BI_1^{\ep}$ correspond to the
light-cone Hamiltonians which are proportional to the
generators $L_0$, $\bar L_0$ of the Virasoro algebra in the
massless cases. For these cases
it can be shown \cite{T} that we have the
following formula for the expectation values of $\BI_n^\ep$
in a state characterized by $p\in\BR$ and tuples $\bk^\ep$:
\begin{equation}\label{lcham}
\BI_{1}^\ep\,=\,\frac{2\pi}{R}\bigg( P^2-\frac{1}{24}+\sum_{a\in \BK}
k^{\ep}_{a}\bigg)\,.
\end{equation}
We clearly identify the zero-mode contribution $\propto p^2$ and
integer-valued oscillator contributions $k^{\ep}_{a}$.
We therefore reproduced already a good part of the expected structure
of the spectrum of the continuum Liouville theory \cite{CT}.

\subsection{Explicit calculation of the reflection amplitude}
\newcommand{\ka}{\kappa}

The reflection amplitude $S_{q_p}$ introduced in Subsection \ref{Sdef}
represents an important piece of data characterizing
Liouville theory. We are now going to explain how to calculate this
quantity for the class of states related to the primary
states of the Liouville conformal field theory. The key observation
underlying this calculation is equation 
\rf{Sfactor} which relates the reflection amplitude to
the asymptotics of the functions $q_p^\ep$
of KdV theory. These asymptotics were found in \cite{T} based on \cite{FL}.
To round off the picture, we will now briefly recall how this works.

Let us first observe, as can be seen e.g. from formula \rf{lcham},
that the states with $M^\ep=0$, $\ep=\pm$, correspond to the
Fock-vacua in the sectors labelled by $p$.
According to
\rf{Sfactor}, we may calculate $R_{p}\equiv S_{q_p}$
if we know the asymptotic behavior of the  q-functions $q_p^\ep(u)$
corresponding to the Fock-vacua. These q-functions $q_p^\ep(u)$
can be characterized as the unique solutions
of the functional equations \rf{CBax}, \rf{CWron}
which have the analytic properties \rf{Canal}, the
asymptotic behavior \rf{Casym}, and the
additional property to be non-vanishing within the strip $\BS_u$.
It was shown in \cite{FL} that
a solution to this set of conditions
is given by the Wronskian of certain solutions
to the ordinary differential equation
\begin{equation}\label{ODE}
\left[-\frac{d^2}{dx^2}-\frac{4}{b^2}p^2+\ka^2\big(e^{2x}+e^{-2x/b^2}\big)
\right]\Psi=0\,.
\end{equation}
This generalizes similar results for other models which
go back to \cite{DT,BLZ1}.
In order to get $q_p^\ep(u)$, consider the
solutions $\Psi_\pm$ to \rf{ODE} which have the
asymptotic behavior
\begin{equation}\begin{aligned}
& \Psi_+\,\sim\,\frac{1}{\sqrt{2\ka}}\exp\left(
\frac{x}{2b^2}-\ka b^2 e^{-x/b^2}\right) \quad
{\rm for}\quad x\ra -\infty\,, \\
& \Psi_-\,\sim\,\frac{1}{\sqrt{2\ka}}\exp\left(-\frac{x}{2}-\ka e^{x}\right)
 \quad
{\rm for}\quad x\ra +\infty\,,
\end{aligned}
\end{equation}
respectively. The functions $q^\ep_p(\vt)$ are then simply
given as
\begin{equation}\label{ODEIM}
q^+_p(u)\,\equiv\,q^-_p(-u)\,\equiv\,
\Psi_+\frac{d}{dx}\Psi_--\Psi_-\frac{d}{dx}\Psi_+\,,
\end{equation}
provided that we identify the respective parameters as
follows,\footnote{Concerning the comparison with \cite{FL}
note that
the parameter $n$ used there is related
to $b^2$ via $n=2/b^2$.}
\begin{equation}
\kappa\,=\,-\frac{\ka_0}{2\sin\vt_0}\,
\frac{mR}{2}\,e^{\frac{\pi}{2\eta}u},
\quad\ka_0\,=\,-
\frac{2\sqrt{\pi}}{\Ga\big(-\frac{1}{2(1+b^2)}\big)
\Ga\big(1-\frac{b^2}{2(1+b^2)}\big)}\,.
\end{equation}
The characterization \rf{ODEIM} of $q^\ep_p(u)$ in terms of the ODE \rf{ODE}
allowed the authors of \cite{FL} to
determine the asymptotics of $q^\ep_p(u)$. The explicit
expression
for $S_p=e^{2i\theta(p)}$ which follows from
formula (177) in \cite{FL}
is 
\begin{equation}\label{Thetadef}
S_p\,=\,-
\rho^{-8i\de p}\,
\frac{\Ga(1+2ibp)\Ga(1+2ib^{-1}p)}{\Ga(1-2ibp)\Ga(1-2ib^{-1}p)}\,,
\end{equation}
in which we have used the abbreviation
\begin{equation}
\label{rhodef}
\rho\,\equiv\, \frac{R}{2\pi}\,\frac{m}{4\sqrt{\pi}}\,
\Ga\bigg(\frac{1}{2+2b^2}\bigg)
\Ga\bigg(1+\frac{b^2}{2+2b^2}\bigg)\,.
\end{equation}
We recover the expression proposed in \cite{ZZ}, for which a full
derivation was given in \cite{TL}. We'd like to stress how different
the present derivation of the reflection amplitude -- based
on the integrable structure of Liouville theory --
is compared to the one in \cite{ZZ,TL}, which was based on the
conformal symmetry.
It would be very interesting further elucidate the interplay
between the integrable and the conformal structure of Liouville theory.

\appendix
\section{Asymptotic behavior of Q-operators}

\setcounter{equation}{0}

Let us first note that the Q-operators for Liouville theory and for the KdV model have the same
asymptotic behavior. To this aim let us consider the eigenvalue equation in the form
\begin{equation}\label{Qeigen}
\langle \,q\,|\,\SQ^\ep_{\rm\sst Liou}(u)\,|\,t\,\rangle\,=\,q^\ep_{}(u)\,\langle\, q\,|\,t\,\rangle\,,
\end{equation}
where $\langle\, q\,|$ is a generalized eigenstate of $\SQ^\ep_{\rm\sst Liou}(u)$ with eigenvalue $q^\ep(u)$, and $|\,t\,\rangle$ is
a test function from a suitable dense subspace $\CT$ of $\CH$ like those defined in \cite{BT}. The left hand side of
\rf{Qeigen} can be represented as
\begin{equation}
\int d\bx'd\bx \;\langle\, q'\,|\,\bx'\,\rangle\,\langle\,\bx'\,|\,\SY^\ep_{\rm\sst Liou}(u)\,|\,\bx\,\rangle\,,
\end{equation}
where $\langle\, q'\,|\equiv \langle \,q\,|\,\SY_{\infty}^{-1}$. 
Following \cite[Section 4.2.]{BT} it
is not hard to see that the bulk of the domain of integration
over $\bx'$, $\bx$ gives contributions which decay exponentially when $|u|\ra\infty$. One may observe, however, that the
integration over $\bx'$ may receive contributions from the region in the integration over $\bx'$ where
$x_r=y_r-\de$, $\de\ra \infty$. This is due to the fact that the wave-function
$\langle\, q'\,|\,\bx'\,\rangle$ has plane-wave like behavior w.r.t. the zero mode $x_0=\sum_{n=1}^{\SRN}x_n$ in this limit.
A look at the formula \rf{Qkern} for the kernel $\langle\,\bx'\,|\,\SY^\ep_{\rm\sst Liou}(u)\,|\,\bx\,\rangle$
then reveals that it becomes equal to the kernel
$\langle\,\bx'\,|\,\SY^\ep_{\rm\sst KdV}(u)\,|\,\bx\,\rangle$ for large $\de$. This observation reduces the
problem to find the asymptotic behavior of $\SQ^\ep_{\rm\sst Liou}(u)$ to the corresponding problem for
$\SQ^\ep_{\rm\sst KdV}(u)$.

To solve this problem, an alternative integral operator
representation will be useful. In order to find it, let us
consider a variant of the Q-operators defined as
\begin{equation}
\tilde\SQ^+(u)\,=\,(\SQ^+(s_+))^{-1}\cdot\SQ^+(u)\,,\qquad
\tilde\SQ^-(u)\,=\,(\SQ^-(s_-))^{-1}\cdot\SQ^-(u)\,.
\end{equation}
One advantage of the Q-operators $\tilde\SQ^+(u)$ and $\tilde\SQ^-(u)$
is the fact that the kernels representing these operators can be written in an even more
explicit form,
\begin{align}\label{Qkern}
\bra  & \,{\mathbf x}'\,|\,\tilde\SQ^+(u)\,|\,{\mathbf x}\,\ket\,=\,
\prod_{n=1}^\SRN W_{2s+i\eta}^{-}(x_{n}'+x_{n+1}')
\overline{W}_{u-s}^{}(x_{n}'-x_r^{})W_{u+s}^{+}(x_{n-1}'+x_n^{})\,,\\
\bra  & \,{\mathbf x}'\,|\,\tilde\SQ^-(u)\,|\,{\mathbf x}\,\ket\,=\,
\prod_{n=1}^\SRN W_{u-s}^-(x_{n-1}'+x_n^{})\overline{W}_{u+s}^{}(x_n'-x_{n}^{})W_{i\eta-2s}^{+}(x_{n}+x_{n+1})
\,,\label{Qbarkern}\end{align}
Let $\langle \,\bt\,|$, $\bt=(t_1,\dots,t_\SRN)$ now be the 
generalized eigenstates of the operators
$\su_n$ such that $\langle\,\bt\,|\,\su_n\,=\,\langle\,\bt\,|\,e^{\pi bt_n}$.
By means of straightforward computations it is possible to show that
\begin{align}\label{t-rep}\nonumber
\langle\,\bt'\,|\,\tilde\SQ_{\rm\sst KdV}^+(u)\,|\,\bt\,\rangle\,=\, 
& \de(p-p')\,E_s \,e^{-\frac{\pi i}{2}\SRN u^2}
\,e^{-2\pi i \tau_r^{}t_r}\\
& \times\int_\BR dx\;e^{4\pi i px}\;\prod_{n=1}^{N}
\vf(w+x+\tau_n)
\vf(w-x-\tau_n)\,,
\end{align}
where $E_s$ is a constant, and we have used the notation 
$2p\equiv\sum_{s=1}^{\SRN}t_s$ and
$\tau_r\equiv\sum_{s=1}^{r-1}(t_s'-t_s^{})$. We are now in the position
to prove that
\begin{equation}
\tilde\SQ_{\rm\sst KdV}^+(u)\underset{\substack{u\ra-\infty\\ 
{\rm Im}(u)={\rm const}}}{\sim}
E_s\, e^{-\frac{\pi i}{2}\SRN u^2}\,\big(\,e^{2\pi i\spp_\0 (u-s)}
\SA^+_++e^{-2\pi i\spp_\0 (u-s)}\SA_-^+\,\big)\,,
\end{equation}
where $\SA^+_\pm$ are operators represented by the kernels
\begin{align}\label{A-t-rep}
\langle\,\bt'\,|\,\SA^+_\pm\,|\,\bt\,\rangle\,=\,& \de(p-p')
\,e^{-2\pi i \tau_r^{}t_r} \int_\BR dy\;e^{\mp 4\pi i py}\;\prod_{r=1}^{N}
\vf\big(y\mp \tau_r+\fr{i}{2}\eta\big)\,,
\end{align}
respectively.
Indeed, it is easy to see that the dominant contributions to the asymptotics
$u\ra\infty$ come from the region in the integration over $x$ where
$|x|\sim u$. In order to isolate the contributions from $x\pm u=\CO(1)$,
respectively, let us change the variable of integration
to $y^\ep=\frac{u-s}{2}
\mp x$. Taking into account that $\vf(x)\sim 1$ for $x\ra\infty$
it becomes easy to verify our claim.

\end{document}